\documentclass[journal]{IEEEtran}

\normalsize

\ifCLASSINFOpdf
 
\else
 
\fi

\usepackage{cite}
\usepackage{amsmath,amssymb,amsfonts}
\usepackage{bm}
\usepackage{algorithm}
\usepackage{algpseudocode}
\usepackage{graphicx}
\usepackage{textcomp}
\usepackage{xcolor}
\usepackage{float}
\usepackage{amsthm}
\usepackage{graphicx}
\usepackage{epstopdf}
\usepackage{amsmath,bm}
\usepackage{amsfonts}
\usepackage{amssymb}
\usepackage{color}
\usepackage{multirow}
\usepackage{multicol}
\usepackage{soul,xcolor}
\usepackage{overpic}
\newtheorem{lemma}{Lemma}

\DeclareMathOperator{\tr}{tr}

\definecolor{orange}{RGB}{0,112,192}
\begin{document}
\makeatletter
\newcommand*{\rom}[1]{\expandafter\@slowromancap\romannumeral #1@}
\makeatother

\title{Joint Power Control and LSFD for Wireless-Powered Cell-Free Massive MIMO 
}

\author{\"Ozlem Tu\u{g}fe Demir,~\IEEEmembership{Member,~IEEE,} and
	Emil Bj\"ornson,~\IEEEmembership{Senior Member,~IEEE}
	\thanks{The authors are with the Department of Electrical Engineering
		(ISY), Linköping University, 581 83 Linköping, Sweden (e-mail: ozlem.tugfe.demir@liu.se, emil.bjornson@liu.se). This work was partially supported by ELLIIT and the Wallenberg AI, Autonomous Systems and Software Program (WASP) funded by the Knut and Alice Wallenberg Foundation. A part of this paper was presented in WCNC 2020 \cite{conf}.}
}

\maketitle

\begin{abstract}
This paper considers wireless uplink information and downlink power transfer in cell-free massive multiple-input multiple-output systems. The single-antenna user equipments (UEs) utilize the energy harvested in the downlink to transmit uplink pilot and information signals to the multiple-antenna access points (APs). We consider Rician fading and maximum ratio processing based on either linear minimum mean-squared error (LMMSE) or least-squares (LS) channel estimation. We derive the average harvested energy by using a practical non-linear energy harvesting circuit model for both coherent and non-coherent transmission schemes. Furthermore, the uplink spectral efficiency (SE) is derived for all the considered methods and the max-min fairness problem is cast where the optimization variables are the AP and UE power control coefficients together with the large-scale fading decoding vectors. The objective is to maximize the minimum SE of the UEs' under APs' and UEs' transmission power constraints. A novel alternating optimization algorithm with guaranteed convergence and improvement at each step is proposed to solve the highly-coupled non-convex problem. 
\end{abstract}
\IEEEpeerreviewmaketitle

\section{Introduction}

Massive MIMO (multiple-input multiple-output) has received great interest in the last decade and has been extensively analyzed for cellular systems due to its high spectral efficiency (SE) achieved by spatial multiplexing of many user equipments (UEs) on the same time-frequency resource  \cite{emil_book,unlimited, erik_book,ozge_massive,massive_mimo_reality, massive_mimo_magazine}. Now, it has reached its mature stage and is one of the key technologies in 5G, and commercial deployments began in 2018 \cite{massive_mimo_reality}. Although 5G cellular technology with massive MIMO is expected to provide higher data rates compared to the previous technologies, the inter-cell interference is still an important issue, particularly for the cell-edge UEs \cite{emil_multiple}. Recently, an alternative network infrastructure is considered in \cite{nayebi,cell_free_vs_small_cell}, which uses the name \emph{cell-free massive MIMO} since a large number of access points (APs) is distributed over a large geographic area to serve all the UEs in a coherent manner without any cell boundaries. Cell-free massive MIMO was shown to improve the minimum SE achieved in the network and total energy efficiency \cite{cell_free_vs_small_cell,cell_free_new1,cell_free_new2}. Recent works have focused on different aspects of cell-free massive MIMO and several network architectures have been proposed \cite{making_cell_free, ozge_cell_free,giovanni,swipt_cell_free,wpt1,wpt2,wpt3,emil_cellfree,emil_scalable}.

Communication and positioning are the main use cases for radio frequency (RF) in current wireless systems. While we are in the era of 5G for mobile communication, some emerging technologies have potential to be integrated into future generation standards. Wireless power transfer (WPT) via RF signals is one of these technologies to exploit the RF energy for battery-limited devices and there has been extensive research conducted in this area to charge mobile battery-powered devices via  ambient and dedicated RF signals \cite{1g,bruno}. WPT would reduce the battery requirements (size, wiring, etc.) of the mobile devices and provide more consistent and ubiquitous service to energy-hungry devices by supplying reliable energy. In particular, future autonomous low-power networks and Internet of Things (IoT) are expected to benefit from this technology \cite{1g}. 

Simultaneous wireless information and power transfer (SWIPT), which is an interesting paradigm in WPT, has been considered for cellular massive MIMO systems \cite{ps_downlink1, ps_downlink2, swipt_downlink, swipt_downlink2}. In these works, the UEs have either a power splitting or time switching circuit to utilize the downlink RF signals for both information reception and energy harvesting. In \cite{wet2,wet_heath}, a base station (BS) assists the UEs for their uplink pilot and data transfer by energy beamforming in the downlink. In this paper, we adopt this setup but consider a cell-free massive MIMO system with several transmission schemes and a practical non-linear energy harvesting model. The motivation behind integrating cell-free massive MIMO with WPT is that each UE is expected to have a much higher channel gain to at least one of the APs with larger probability compared to cellular massive MIMO. Hence, one of the main limitations in RF WPT, which is path loss, is overcome to some extent. Furthermore, not only sensor networks and IoT devices for which changing batteries is infeasible, but also for the battery-limited mobile UEs, WPT is a promising technology as long as the transmission range is not too long \cite{1g, wet2, wet_heath,wpt_mobile}. Hence, cell-free networks are advantageous compared to cellular systems in this respect. In this paper, we assume all the UEs in the network benefit from WPT and as the simulations show, employing denser APs improves the minimum guaranteed uplink SE. 

The prior works on WPT in cell-free systems are few. In \cite{wpt1}, the total harvested energy throughput is maximized together with the AP selection under transmission power constraints for each AP. This work assumes perfect channel state information (CSI) and does not take into account the uplink communications. In \cite{swipt_cell_free}, SWIPT is considered in the context of cell-free massive MIMO where information and energy UEs are located separately. Similarly, \cite{wpt2} studied cell-free massive MIMO where the information UEs do not harvest energy and there is a single energy-harvesting UE that actively eavesdrops. In \cite{wpt3}, the authors consider minimization of the total transmitted energy for wirelessly-powered cell-free IoT by considering only Rayleigh fading and a linear energy harvesting model. In this paper, we adopt a more general Rician fading channel model with a common random phase shift to all the antennas at each AP. This leads to the channel coefficients for different AP antennas being dependent on each other. Furthermore, the channels are not Gaussian unlike most of the previous works. Hence, the statistical results presented in previous works cannot be used here. We derive the exact closed-form expressions for SE and average harvested energy for the first time for Rician fading with random phase shifts by using two different channel estimation schemes.

 Different from the existing works, this paper is the first one that considers power control for maximizing the minimum uplink SE for downlink WPT-assisted cell-free massive MIMO. Max-min fairness is one of the important optimization criteria since it maximizes the minimum guaranteed SE to all the UEs, which is highly in accordance with the uniformly great service goal of cell-free systems. Furthermore, max-min fairness may be effective to reduce the traffic congestion mainly resulting from the UEs in bad channel conditions, by increasing the \%95-likely SE of the whole network. The resulting joint optimization problem in terms of the uplink/downlink power coefficients and large-scale fading decoding (LSFD) weights is non-convex and more challenging compared to the previous works due to the non-linear energy harvesting model and highly-coupled variables. After some mathematical manipulations, we come up with a problem structure where an efficient modified bisection search-based alternating optimization can be applied. We show that semidefinite programming with rank relaxation guarantees a rank one solution for the subproblems. Overall, the main contributions of this paper are: 
\begin{itemize}
	\item We derive the average harvested energy and the uplink SE in closed-form when the channels are estimated using a linear minimum mean-squared error (LMMSE) and least squares (LS) estimators for practical Rician fading channels with unknown phase shifts. We derive the SE expressions for the multi-antenna APs that are generalizations of the SE for single-antenna APs in \cite{ozge_cell_free}. Note that the results in \cite{ozge_cell_free} cannot be used for multiple-antenna APs due to the common phase shifts of the channels, which are the same for each BS antenna. 
	
	\item We consider both coherent and non-coherent downlink WPT schemes where the same or independent energy symbol for each UE is transmitted from the APs, respectively. Furthermore, a practical non-linear energy harvesting model \cite{revise1} is utilized in the closed-form results and the optimization algorithm.
	
	\item We formulate the max-min fair joint AP and UE power control and LSFD design problem under the constraints on harvested and transmitted power at the APs and UEs. Note that this problem has a different structure than \cite{wpt3}, which does not consider LSFD, non-linear energy harvesting, or max-min fairness. Furthermore, we consider a different pilot signaling scheme than the random pilot signaling \cite{wpt3}. These factors make our problem unique and more challenging compared to the case with a linear energy harvesting model and no LSFD.
	
	\item We propose an alternating optimization algorithm to achieve a convergent solution to the proposed non-convex problem. The resulting non-convex sub-problems are solved efficiently after some novel transformations. The simulation results show that the solution found by this algorithm improves the minimum guaranteed SE of the network compared to simpler power control scheme in \cite{giovanni}. 
\end{itemize}
  
Note that the conference version of this paper, \cite{conf}, only considers LMMSE-based channel estimation and non-coherent energy transmission using a linear energy harvesting model.

{\bf Reproducible research:} All the simulation results can be
reproduced using the Matlab code and data files available at:
https://github.com/emilbjornson/wireless-powered-cell-free

\section{System Model}

We consider a cell-free massive MIMO system where $L$ multiple-antenna APs are distributed over a large area to serve $K$ single-antenna UEs with energy harvesting capability. Each AP is equipped with $N$ antennas and connected to a central processing unit (CPU) via an error-free fronthaul link. In this paper, we assume time division duplex (TDD) operation and, hence, channel reciprocity holds. Let $\tau_c$ denote the total number of samples per coherence interval. Each coherence interval is divided into three phases: uplink training, downlink WPT, and uplink wireless information transfer (WIT). In the uplink training phase, all the UEs send pilot sequences of length $\tau_p$ to the APs, which estimate the channels to design precoding vectors for effective energy transfer and data reception. While $\tau_d$ samples are used for downlink WPT, the remaining $\tau_u$ samples are used for the uplink WIT, hence, we have $\tau_p+\tau_d+\tau_u=\tau_c$. In accordance with the existing literature on cell-free massive MIMO, the CSI is not shared between the APs \cite{cell_free_vs_small_cell}, \cite{making_cell_free}.

Let ${\bf g}_{kl}\in \mathbb{C}^{N}$  denote the channel between the $k^{\textrm{th}}$ UE and the $l^\textrm{th}$ AP. The channels are constant in each time-frequency coherence interval. We consider spatially uncorrelated Rician fading channels with unknown phase shifts, which is the first novelty of this paper in the context of cell-free massive MIMO with multiple-antenna APs. This means each channel realization can be expressed as
\begin{align} \label{eq:channel}
& {\bf g}_{kl}=e^{j\theta_{kl}}{\bf \bar{g}}_{kl}+\tilde{\bf g}_{kl},
\end{align}
where $e^{j\theta_{kl}}{\bf \bar{g}}_{kl}\in \mathbb{C}^{N}$  denotes the line-of-sight (LOS) component. Moreover, $\tilde{\bf g}_{kl}$ is the non-line-of-sight (NLOS) component and the small-scale fading is modeled as $\tilde{\bf g}_{kl} \sim \mathcal{N}_{\mathbb{C}}({\bf 0}_N,\beta_{kl}{\bf I}_N)$, where $\beta_{kl}$ is the large-scale fading coefficient which accounts for  path-loss and shadowing. Note that the vectors $\{{\bf \bar{g}}_{kl}\}$ and large-scale fading coefficients $\{\beta_{kl}\}$ describe the long-term channel effects and change more slowly than the small-scale fading realizations. We assume that the APs have perfect knowledge of $\{{\bf \bar{g}}_{kl}, \beta_{kl}\}$ corresponding to the channels between them and the UEs, in accordance with prior literature \cite{emil_book}, \cite{ozge_massive}. We consider a realistic scenario where the phase shifts $\{\theta_{kl}\}$ in the LOS components are unknown due to user mobility  and assume they are uniformly distributed in the interval $[0,2\pi)$ \cite{ozge_cell_free}. Note that most of the previous works that consider Rician fading neglect the phase shifts $\{\theta_{kl}\}$ since they do not affect the distribution of the channel gain. However, in practical systems where channel estimation is required at the receiver, we should take these random phase shifts into account. When the transmitter and receiver move over distances at order of the wavelength, a small random phase shift is induced on the LOS component as well as the individual paths constructing the NLOS component of a channel. The effect of the random phase shifts on the large number of scattered paths is already taken into account by modeling the NLOS component of the channel as Gaussian. However, the direct path leading to the LOS component is usually much stronger than the NLOS part and we should treat its phase shift separately. These phase shifts vary at the same pace as the NLOS component and from coherence block to coherence block. As a result, the phase shifts are not known in advance at the BS as NLOS components and we should consider the unknown phase shifts on the direct path separately in channel estimation.

\section{Channel Estimation}

Let $\bm{\varphi}_k \in \mathbb{C}^{\tau_p}$ denote the pilot sequence that is assigned to the $k^\textrm{th}$ UE where $||\bm{\varphi}_{k}||^2=\tau_p$. If $\tau_p\geq K$, we can use an orthogonal set of pilot sequences. However, this is generally not the case for cell-free massive MIMO systems since the number of UEs can be much larger than the pilot sequence length. Hence, so-called pilot contamination occurs. The second novelty of this paper is to take pilot contamination into account in the analysis of cell-free massive MIMO based WPT.

Deriving the MMSE estimator is non-trivial since we do not have a linear Gaussian signal model. We will therefore restrict ourselves to the LMMSE estimator as in \cite{ozge_cell_free}, which is the conventional benchmark in the massive MIMO literature. To obtain the LMMSE channel estimator in a simple form, let us assume that the pilot sequences are either identical or mutually orthogonal and call $\mathcal{P}_k$ the subset of UEs which are assigned the same pilot sequence as the $k^\textrm{th}$ UE, including itself.  Then, the received pilot signal ${\bf Z}_l\in \mathbb{C}^{N\times \tau_p}$ at the $l^\textrm{th}$ AP is given by
\begin{align}\label{eq:pilot}
& {\bf Z}_{l}=\sum_{k=1}^{K}\sqrt{\rho_p}{\bf g}_{kl}\bm{\varphi}_{k}^T+{\bf N}_l,
\end{align}
where $\rho_p$ is the pilot transmit power and the additive noise matrix ${\bf N}_{l}\in \mathbb{C}^{N \times \tau_p}$ has i.i.d. $\mathcal{N}_{\mathbb{C}}(0,\sigma^2)$ random variables. A sufficient statistics for the estimation of the $k^\textrm{th}$ UE's channel is  
\begin{align}\label{eq:suff-stats} 
& {\bf z}_{kl}=\frac{{\bf Z}_l\bm{\varphi}_k^{*}}{\sqrt{\tau_p}}=\sqrt{\tau_p\rho_p}\sum_{i \in \mathcal{P}_k}{\bf g}_{il}+{\bf n}_{kl},
\end{align}
where ${\bf n}_{kl}\triangleq{\bf N}_l\bm{\varphi}_k^{*}/\sqrt{\tau_p}\sim \mathcal{N}_{\mathbb{C}}({\bf 0}_N,\sigma^2{\bf I}_N)$. Note that ${\bf n}_{il}$ is independent of ${\bf n}_{kl}$ for $\forall i \notin \mathcal{P}_k$. Then, the phase-unaware LMMSE estimate  of ${\bf g}_{kl}$, based on \eqref{eq:suff-stats}, is 
\begin{align}\label{eq:lmmse} 
&{\bf \hat{g}}_{kl}=\sqrt{\tau_p\rho_p}{\bf R}_{kl}{\bf \Psi}_{kl}^{-1}{\bf z}_{kl},
\end{align}
where 
\begin{align} 
&{\bf R}_{kl}\triangleq\mathbb{E}\{{\bf g}_{kl}{\bf g}_{kl}^H\}={\bf \bar{g}}_{kl}{\bf \bar{g}}_{kl}^H+\beta_{kl}{\bf I}_N  \label{eq:Rkl},\\
&{\bf \Psi}_{kl}\triangleq\mathbb{E}\{{\bf z}_{kl}{\bf z}_{kl}^H\}=\tau_p\rho_p\sum_{i \in \mathcal{P}_k}\left({\bf \bar{g}}_{il}{\bf \bar{g}}_{il}^H+\beta_{il}{\bf I}_N\right)+\sigma^2{\bf I}_N. \label{eq:Psi}
\end{align}
The channel estimate ${\bf \hat{g}}_{kl}$ and the estimation error ${\bf e}_{kl}={\bf g}_{kl}-\hat{\bf g}_{kl}$ are zero-mean uncorrelated random vectors with covariance matrices 
\begin{align}
&{\bf \hat{R}}_{kl}\triangleq\mathbb{E}\left\{{\bf \hat{g}}_{kl}{\bf \hat{g}}_{kl}^H\right\}=\tau_p\rho_p{\bf {R}}_{kl}{\bf {\Psi}}_{kl}^{-1}{\bf {R}}_{kl}, \label{eq:Rhat} \\
&{\bf C}_{kl}\triangleq\mathbb{E}\left\{{\bf e}_{kl}{\bf e}_{kl}^H\right\}={\bf {R}}_{kl}-\tau_p\rho_p{\bf {R}}_{kl}{\bf {\Psi}}_{kl}^{-1}{\bf {R}}_{kl}.
\end{align}
Note that neither the channel estimate nor the estimation error is Gaussian. As a result, although they are uncorrelated, they are not independent.

Note that the LMMSE-based channel estimator presented above requires the computation of an inverse matrix which can be computationally complex when $N$ is large. A simpler estimation scheme is the  least squares (LS) estimator that does not make use of the channel statistics. The LS estimate of the channel ${\bf g}_{kl}$ is a scaled version of ${\bf z}_{kl}$ in \eqref{eq:suff-stats}. In the following parts of the paper, we will use directly ${\bf z}_{kl}$ for maximum ratio (MR) processing since power control optimization will be implemented and, hence, the scaling factor in front of ${\bf z}_{kl}$ will not affect the result.

\section{Downlink Energy Harvesting}

In the WPT phase, each AP transmits energy to the UEs by using the CSI for downlink precoding. In this paper, we will first analyze  coherent energy transmission where the APs transmit the same energy symbol for each UE in a synchronous manner in order to increase the harvested energy at the UEs.

Let ${\bf w}_{kl}^{*}\in \mathbb{C}^{N}$ denote the downlink precoding vector for the energy harvesting phase. Then, the signal transmitted by the $l^\textrm{th}$ AP is 
\begin{align} \label{eq:tr_energy}
&{\bf x}_l^E=\sum_{k=1}^K\sqrt{p_{kl}}{\bf w}_{kl}^{*}s_{k},
\end{align}
where $s_{k}$  is the zero-mean unit-variance energy signal for the $k^{\textrm{th}}$ UE. The energy signals for different UEs are assumed independent for the ease of analysis. $p_{kl}$ is the power control coefficient of the $l^{\textrm{th}}$ AP corresponding to the $k^{\textrm{th}}$ UE. The transmission power for each AP should satisfy the maximum power limit that is $\rho_d$ in the long-term, i.e.,
\begin{align} \label{eq:max_power}
&P_{l}^E\triangleq\mathbb{E}\left\{\left\Vert{\bf x}_l^E\right\Vert^2\right\}\leq \rho_d.
\end{align}
The average transmitted power $P_l^E$ for the $l^{\textrm{th}}$ AP is
\begin{align} \label{eq:Pl}
&P_{l}^E=\mathbb{E}\left\{\left\Vert\sum_{k=1}^K\sqrt{p_{kl}}{\bf w}_{kl}^{*}s_{k}\right\Vert^2\right\}\stackrel{(a)}=\sum_{k=1}^Kp_{kl}\mathbb{E}\left\{\left\Vert{\bf w}_{kl}\right\Vert^2\right\}\nonumber\\
\stackrel{(b)}=&\begin{cases}\sum\limits_{k=1}^Kp_{kl}\tr\left({\bf \hat{R}}_{kl}\right), & \text{if LMMSE with } {\bf w}_{kl}={\bf \hat{g}}_{kl} \text{ in } \eqref{eq:lmmse}, \\
\sum\limits_{k=1}^Kp_{kl}\tr\left({\bf \Psi}_{kl}\right), & \text{if LS with } {\bf w}_{kl}={\bf z}_{kl} \text{ in } \eqref{eq:suff-stats},
\end{cases}
\end{align}
where $(a)$ is the result of the independence of the zero-mean signals $\left\{s_{k}\right\}$. We used \eqref{eq:Psi} and \eqref{eq:Rhat} in $(b)$ for the MR precoders based on the LS and LMMSE-based channel estimation.
The received signal in the energy harvesting phase at the $k^{\textrm{th}}$ UE is 
\begin{align}\label{eq:rek}
r_k^E&=\sum_{l=1}^L{\bf g}_{kl}^T{\bf x}_l^E+n_k^E=\sum_{l=1}^L\sum_{i=1}^K\sqrt{p_{il}}{\bf w}_{il}^H{\bf g}_{kl}s_{i}+n_k^E,
\end{align}
where $n_k^E \sim \mathcal{N}_{\mathbb{C}}(0,\sigma^2)$ is the additive noise at the $k^{\textrm{th}}$ UE. Since the noise floor is too low for energy harvesting, we simply neglect the effect of $n_k^E$ in the average harvested energy expression in accordance with the existing literature \cite{wet_heath,swipt_cell_free,wpt1}. Then, the average input power at the energy harvesting rectifier circuit of the $k^\textrm{th}$ UE  is
\begin{align} \label{eq:Ik}
I_{k}=&\mathbb{E}\left\{\left|\sum_{l=1}^L\sum_{i=1}^K\sqrt{p_{il}}{\bf w}_{il}^H{\bf g}_{kl}s_{i}\right|^2\right\}\nonumber\\\stackrel{(a)}=&\sum_{l=1}^L\sum_{l^{\prime}=1}^L\sum_{i=1}^K\sqrt{p_{il}}\sqrt{p_{il^{\prime}}}\mathbb{E}\left\{{\bf w}_{il}^H{\bf g}_{kl}{\bf g}_{kl^{\prime}}^H{\bf w}_{il^{\prime}}\right\} \nonumber\\
\stackrel{(b)}=&\sum_{l=1}^L\sum_{i=1}^Kp_{il}\mathbb{E}\left\{{\bf w}_{il}^H{\bf g}_{kl}{\bf g}_{kl}^H{\bf w}_{il}\right\} \nonumber\\&+\sum_{l=1}^L\sum_{\substack{l^{\prime}=1, \\ l^{\prime}\neq l}}^L\sum_{i=1}^K\sqrt{p_{il}}\sqrt{p_{il^{\prime}}}\mathbb{E}\left\{{\bf w}_{il}^H{\bf g}_{kl}\right\}\mathbb{E}\left\{{\bf g}_{kl^{\prime}}^H{\bf w}_{il^{\prime}}\right\} 
\end{align}
where $(a)$ and $(b)$ follow from the independence of the zero-mean energy signals and the channels to different APs.

The following lemma presents the average input power to the harvester, $I_k$, analytically for coherent energy transmission with the two MR precoders.

\begin{lemma}
	The average input power at the energy harvesting circuit of the $k^{th}$ UE for coherent energy transmission is given by
	\begin{align} 
	I_k&=\sum_{l=1}^L\sum_{i=1}^Kp_{il}\tr\left({\bf \hat{R}}_{il}{\bf R}_{kl}\right)+\sum_{l=1}^L\sum_{i\in \mathcal{P}_k}p_{il}\tau_p^2\rho_p^2\times\nonumber\\
	&\Bigg(2\beta_{kl}\Re\left\{{\bf \bar{g}}_{kl}^H{\bf \Psi}_{il}^{-1}{\bf R}_{il}{\bf \bar{g}}_{kl}\tr\left({\bf R}_{il}{\bf \Psi}_{il}^{-1}\right)\right\}\nonumber\\&+\beta_{kl}^2\left|\tr\left({\bf \Psi}_{il}^{-1}{\bf R}_{il}\right)\right|^2\Bigg) +\sum_{l=1}^L\sum_{\substack{l^{\prime}=1, \\ l^{\prime}\neq l}}^L\sum_{i\in \mathcal{P}_K}\sqrt{p_{il}}\sqrt{p_{il^{\prime}}}\tau_p^2\rho_p^2\times\nonumber\\
	&\Bigg({\bf \bar{g}}_{kl}^H{\bf \Psi}_{il}^{-1}{\bf R}_{il}{\bf \bar{g}}_{kl}+\beta_{kl}\tr\left({\bf \Psi}_{il}^{-1}{\bf R}_{il}\right)\Bigg)\times\nonumber\\
	&\Bigg({\bf \bar{g}}_{kl^{\prime}}^H{\bf \Psi}_{il^{\prime}}^{-1}{\bf R}_{il^{\prime}}{\bf \bar{g}}_{kl^{\prime}}+\beta_{kl^{\prime}}\tr\left({\bf \Psi}_{il^{\prime}}^{-1}{\bf R}_{il^{\prime}}\right)\Bigg)^*    \text{with LMMSE}, \label{eq:Ik-lemma1} 
	\end{align}
	\begin{align}
	I_k&=\sum_{l=1}^L\sum_{i=1}^Kp_{il}\tr\left({\bf \Psi}_{il}{\bf R}_{kl}\right)+\sum_{l=1}^L\sum_{i\in \mathcal{P}_k}p_{il}\tau_p\rho_p\times\nonumber\\&\Big(2N\beta_{kl}{\bf \bar{g}}_{kl}^H{\bf \bar{g}}_{kl}+N^2\beta_{kl}^2\Big) +\sum_{l=1}^L\sum_{\substack{l^{\prime}=1, \\ l^{\prime}\neq l}}^L\sum_{i\in \mathcal{P}_K}\sqrt{p_{il}}\sqrt{p_{il^{\prime}}}\tau_p\rho_p\times\nonumber\\&\Big({\bf \bar{g}}_{kl}^H{\bf \bar{g}}_{kl}+N\beta_{kl}\Big)\Big({\bf \bar{g}}_{kl^{\prime}}^H{\bf \bar{g}}_{kl^{\prime}}+N\beta_{kl^{\prime}}\Big)^* \ \ \text{with LS}. 	 \label{eq:Ik-lemma2}
	\end{align}

	\begin{IEEEproof} Please see Appendix~\ref{lemma1_proof}.
		\end{IEEEproof}
\end{lemma}

Note that all the terms in \eqref{eq:Ik-lemma1} and \eqref{eq:Ik-lemma2} are positive. All UEs' intended signals from all the APs make a contribution to the input power of the rectifier circuit for each UE. The value is also affected by the power control coefficients $\{p_{il}\}$. In addition to the first summation, having pilot contaminated channel estimates brings some additional power terms into the second summation. However, at the same time, the pilot contamination reduces the channel estimation quality, which leads to a reduction in some of the expressions related to correlation between the actual and estimated channels. Hence, it is not easily seen from the formula whether the pilot contamination increases the harvested energy or not. As expected, the input power for energy harvesting increases with the increase in the large-scale fading coefficients $\{\beta_{kl}\}$ and the norm of the LOS parts of the channels $\{\Vert\bar{\bf g}_{kl}\Vert\}$. From \eqref{eq:Ik-lemma2}, it is clearly seen that using larger number of antennas, $N$, at the APs increase the input power of the rectifier.

In the prior conference version of this work, non-coherent energy transmission that allows each AP to transmit their choice of energy symbols is analyzed. Hence, the non-coherent scheme does not require any synchronization among APs in the downlink. For this case, the average input power at the energy harvesting circuit of the $k^{\textrm{th}}$ UE is given by
\begin{align} \label{eq:Ik_nonhoh}
I_{k}=&\sum_{l=1}^L\sum_{i=1}^Kp_{il}\mathbb{E}\left\{\left|{\bf w}_{il}^H{\bf g}_{kl}\right|^2\right\}, 
\end{align}
which is the first term of $I_k$ for the coherent energy transmission in \eqref{eq:Ik}. Hence, using Lemma~1, the average input power $I_k$ for the non-coherent energy transmission case is given by
	\begin{align} \label{eq:Ik_noncoh-lemma1}
I_k&=\sum_{l=1}^L\sum_{i=1}^Kp_{il}\tr\left({\bf \hat{R}}_{il}{\bf R}_{kl}\right)+\sum_{l=1}^L\sum_{i\in \mathcal{P}_k}p_{il}\tau_p^2\rho_p^2\times\nonumber\\&\Bigg(2\beta_{kl}\Re\left\{{\bf \bar{g}}_{kl}^H{\bf \Psi}_{il}^{-1}{\bf R}_{il}{\bf \bar{g}}_{kl}\tr\left({\bf R}_{il}{\bf \Psi}_{il}^{-1}\right)\right\}\nonumber\\
&+\beta_{kl}^2\left|\tr\left({\bf \Psi}_{il}^{-1}{\bf R}_{il}\right)\right|^2\Bigg)   \ \ \ \ \text{with LMMSE}, \\
\label{eq:Ik_noncoh-lemma2}
I_k&=\sum_{l=1}^L\sum_{i=1}^Kp_{il}\tr\left({\bf \Psi}_{il}{\bf R}_{kl}\right)\nonumber\\&+\sum_{l=1}^L\sum_{i\in \mathcal{P}_k}p_{il}\tau_p\rho_p\Big(2N\beta_{kl}{\bf \bar{g}}_{kl}^H{\bf \bar{g}}_{kl}+N^2\beta_{kl}^2\Big)  \ \  \ \text{with LS}.
\end{align}
The average input power for the non-coherent transmission is always less than that of the coherent transmission. Hence, on the average, the coherent scheme allows more energy harvesting at each UE. In the simulation results, we will quantify this benefit.

We will use the following non-linear energy harvesting model in accordance with \cite{revise1}. This model highly correlates with real measured data. The total harvested energy at the $k^{\textrm{th}}$ UE in $\tau_d$ channel uses is
\begin{align}
E_k=\frac{\tau_dA_kI_k}{B_kI_k+C_k} \label{eq:Ek}
\end{align} 
where $A_k>0$, $B_k\geq0$, and $C_k$ are constants determined by curve fitting of the rectifier circuit of the $k^{\textrm{th}}$ UE \cite{revise1}. If we set $B_k$ to zero, we obtain the classical linear energy harvesting model. 

Note that, the harvested energy is proportional to the number of downlink energy symbols, $\tau_d$. However, increasing $\tau_d$ is expected to increase the SE up to some extent since for a fixed coherence block length, $\tau_c$, an increase in $\tau_d$ necessitates a decrease in $\tau_u$ that is proportional to the SE of each UE as we consider in the next section.

\section{Uplink Wireless Information Transfer}

In the uplink information transmission phase, all the UEs simultaneously send their data signals to the APs. Let $q_k$ denote the symbol of the $k^{\textrm{th}}$ UE, which is zero-mean with $\mathbb{E}\left\{|q_k|^2\right\}=1$, and $\eta_k\geq0$ is the corresponding transmission power. The received signal at the $l^{\textrm{th}}$ AP is
\begin{align} \label{eq:received_AP}
{\bf r}_l^I=\sum_{k=1}^K\sqrt{\eta_k}{\bf g}_{kl}q_k+{\bf n}_l^I, \ \ l=1,\ldots,L,
\end{align}
where ${\bf n}_l^I \sim \mathcal{N}_{\mathbb{C}}\left({\bf 0}_N,\sigma^2{\bf I}_N\right)$ is the additive white Gaussian noise. Each AP applies local decoding for each UE's information symbol before sending it to the CPU. Let ${\bf v}_{kl}^* \in \mathbb{C}^{N}$ denote the decoding weight vector for the $k^{\textrm{th}}$ UE's signal at the $l^{\textrm{th}}$ AP. Hence, $\tilde{r}_{kl}={\bf v}_{kl}^H{\bf r}_l^I$ is the locally decoded signal for the $k^{\textrm{th}}$ UE at the $l^{\textrm{th}}$ AP. We consider two MR decoding methods based on the LMMSE or LS-based channel estimates, i.e., ${\bf v}_{kl}={\bf \hat{g}}_{kl}$ or ${\bf v}_{kl}={\bf z}_{kl}$, respectively. 

The CPU computes a weighted sum of the locally decoded signals using the large-scale fading decoding (LSFD) method \cite{nayebi}:
\begin{align} \label{eq:LSFP_CPU}
\hat{q}_k=\sum_{l=1}^L a_{kl}^*\tilde{r}_{kl}, 
\end{align}
for the detection of the $k^{\textrm{th}}$ UE's information signal where $\{a_{kl}^*\}$ are the LSFD weights. We assume the CPU uses only the statistical knowledge of the channels so that no CSI sharing is needed  \cite{nayebi,ozge_cell_free,making_cell_free}. Using the SE analysis technique in \cite{erik_book}, we can express $\hat{q}_k$ for the $k^{\textrm{th}}$ UE data detection as
\begin{align}\label{eq:received_CPU2}
\hat{q}_k=\text{DS}_kq_k+\text{BU}_kq_k+\sum_{i\neq k}\text{UI}_{ki}q_{i}+\tilde{n}_k,
\end{align}
where $\text{DS}_k$, $\text{BU}_k$, $\text{UI}_{ki}$ denote the strengths of the desired signal (DS), beamforming gain uncertainty (BU), and the interference of the ${i}^{\textrm{th}}$ UE on the $k^{\textrm{th}}$ UE, while $\tilde{n}_k$ is the total noise at the CPU. $\text{DS}_k$,  $\text{BU}_k$, $\text{UI}_{ki}$, and $\tilde{n}_k$ are given by
\vspace{-2mm}
\begin{align}
& \text{DS}_k=\sqrt{\eta_k}\sum_{l=1}^La_{kl}^*\mathbb{E}\left\{ {\bf v}_{kl}^H{\bf g}_{kl} \right\}, \nonumber\\& \text{BU}_k=\sqrt{\eta_k}\sum_{l=1}^La_{kl}^*\left({\bf v}_{kl}^H{\bf g}_{kl}-\mathbb{E}\left\{ {\bf v}_{kl}^H{\bf g}_{kl} \right\}\right), \label{eq:BU}\\
& \text{UI}_{ki}=\sqrt{\eta_{i}}\sum_{l=1}^La_{kl}^*{\bf v}_{kl}^H{\bf g}_{il}, \quad\quad \tilde{n}_k=\sum_{l=1}^La_{kl}^*{\bf v}_{kl}^H{\bf n}_l^I. \label{eq:nk}
\end{align}
Let us define the following vectors and matrices for ease of notation:
\vspace{-2mm}
\begin{align}
& \bm{a}_k \triangleq[ \ a_{k1} \ \ldots \ a_{kL} \ ]^T\in \mathbb{C}^{L}, \nonumber\\& \bm{b}_k \triangleq [ \ b_{k1} \ \ldots \ b_{kL} \ ]^T\in \mathbb{C}^{L}, \ \ b_{kl}\triangleq \mathbb{E}\left\{ {\bf v}_{kl}^H{\bf g}_{kl}\right\} \label{bk}, \\
&\bm{C}_{ki}\in \mathbb{C}^{L \times L}, \ \ c_{ki}^{ll^{\prime}}\triangleq\mathbb{E}\left\{ {\bf v}_{kl}^H{\bf g}_{il}{\bf g}_{il^{\prime}}^H{\bf v}_{kl^{\prime}}\right\}, \nonumber\\&\bm{D}_{k}\in \mathbb{C}^{L \times L}, \ \ d_{kl}\triangleq\mathbb{E}\left\{ {\bf v}_{kl}^H{\bf n}_l^I\left({\bf n}_l^I\right)^H{\bf v}_{kl}\right\}, \label{dk} 
\end{align}
where $c_{ki}^{ll^{\prime}}$ is the $(l,l^{\prime})$th element of the matrix $\bm{C}_{ki}$. $\bm{D}_k$ is a diagonal matrix with the $l^{\textrm{th}}$ diagonal element being $d_{kl}$.

Utilizing the use-and-then-forget capacity bounding technique in \cite{erik_book}, the uplink  SE for the $k^{\textrm{th}}$ UE with LSFD for any finite value of $L,K,$ and $N$ is given by
\begin{align}
&R_k=\frac{\tau_u}{\tau_c}\log_2 \left(1+\text{SINR}_k\right), \label{eq:rk} 
\end{align}
where the effective signal-to-noise-plus-ratio is
\begin{align}
\text{SINR}_k=\frac{\eta_k\left|\bm{a}_k^H\bm{b}_k\right|^2}{\bm{a}_k^H\left(\sum_{i=1}^K\eta_{i}\bm{C}_{ki}\right)\bm{a}_k-\eta_k\left|\bm{a}_k^H\bm{b}_k\right|^2+\bm{a}_k^H\bm{D}_k\bm{a}_k} \label{eq:sinr}.
\end{align}
In the following two lemmas, we present the uplink SE for the two MR-based decoding vectors, which is another novelty of this paper in the context of multiple antenna cell-free massive MIMO with unknown phase-shifted Rician fading and LSFD.

\begin{lemma} The uplink  SE for the $k^{\textrm{th}}$ UE with MR decoding ${\bf v}_{kl}={\bf \hat{g}}_{kl}$ in \eqref{eq:lmmse} is given in \eqref{eq:rk} with the effective SINR as in \eqref{eq:sinr}, where the elements of $\bm{b}_k$, $\bm{C}_{ki}$, and $\bm{D}_k$ are given as
\begin{align}
 b_{kl}&=\tau_p\rho_p{\bf \bar{g}}_{kl}^H{\bf \Psi}_{kl}^{-1}{\bf R}_{kl}{\bf \bar{g}}_{kl}+\tau_p\rho_p\beta_{kl}\tr\left({\bf \Psi}_{kl}^{-1}{\bf R}_{kl}\right) , \label{eq:bkl} \\
c_{ki}^{ll}&=\tr\left({\bf \hat{R}}_{kl}{\bf R}_{il}\right)+\mathcal{I}_{i\in \mathcal{P}_k}\tau_p^2\rho_p^2\times\nonumber\\&\Big(2\beta_{il}\Re\left\{{\bf \bar{g}}_{il}^H{\bf \Psi}_{kl}^{-1}{\bf R}_{kl}{\bf \bar{g}}_{il}\tr\left({\bf R}_{kl}{\bf \Psi}_{kl}^{-1}\right)\right\}\nonumber\\&+\beta_{il}^2\left|\tr\left({\bf \Psi}_{kl}^{-1}{\bf R}_{kl}\right)\right|^2\Big), \\
c_{ki}^{ll^{\prime}}&=\mathcal{I}_{i\in \mathcal{P}_k}\tau_p^2\rho_p^2\Big({\bf \bar{g}}_{il}^H{\bf \Psi}_{kl}^{-1}{\bf R}_{kl}{\bf \bar{g}}_{il}+\beta_{il}\tr\left({\bf \Psi}_{kl}^{-1}{\bf R}_{kl}\right)\Big)\times\nonumber\\&\Big({\bf \bar{g}}_{il^{\prime}}^H{\bf \Psi}_{kl^{\prime}}^{-1}{\bf R}_{kl^{\prime}}{\bf \bar{g}}_{il^{\prime}}+\beta_{il^{\prime}}\tr\left({\bf \Psi}_{kl^{\prime}}^{-1}{\bf R}_{kl^{\prime}}\right)\Big)^*,  \ \ \ l^{\prime}\neq l, \\
 d_{kl}&=\sigma^2\tr\left({\bf \hat{R}}_{kl}\right), \label{eq:dk}
\end{align}
where $\mathcal{I}_{(.)}$ is the indicator function, i.e., $\mathcal{I}_{i\in\mathcal{P}_k}$ is equal to one if $i \in \mathcal{P}_{k}$, otherwise it is equal to zero.

\begin{IEEEproof}
	Please see the Appendix~\ref{lemma2_proof}.
\end{IEEEproof}
\end{lemma}

\begin{lemma} The uplink  SE for the $k^{\textrm{th}}$ UE with MR decoding ${\bf v}_{kl}={\bf z}_{kl}$ in \eqref{eq:suff-stats} is given in \eqref{eq:rk} with the effective SINR as in \eqref{eq:sinr}, where the elements of $\bm{b}_k$, $\bm{C}_{ki}$, and $\bm{D}_k$ are given as
	\begin{align}
	& b_{kl}=\sqrt{\tau_p\rho_p}{\bf \bar{g}}_{kl}^H{\bf \bar{g}}_{kl}+N\sqrt{\tau_p\rho_p}\beta_{kl}, \label{eq:bkl2} \\
	&c_{ki}^{ll}=\tr\left({\bf \Psi}_{kl}{\bf R}_{il}\right)+\mathcal{I}_{i\in \mathcal{P}_k}\tau_p\rho_p\Big(2N\beta_{il}{\bf \bar{g}}_{il}^H{\bf \bar{g}}_{il}+N^2\beta_{il}^2\Big), \\
	&c_{ki}^{ll^{\prime}}=\mathcal{I}_{i\in \mathcal{P}_k}\tau_p\rho_p\big({\bf \bar{g}}_{il}^H{\bf \bar{g}}_{il}+N\beta_{il}\big)\big({\bf \bar{g}}_{il^{\prime}}^H{\bf \bar{g}}_{il^{\prime}}+N\beta_{il^{\prime}}\big), \ \ \ l^{\prime}\neq l, \\
	& d_{kl}=\sigma^2\tr\left({\bf \Psi}_{kl}\right). \label{eq:dk2}
	\end{align}

	\begin{IEEEproof}
	The proof follows similar steps in the Appendix~\ref{lemma2_proof} and hence omitted.
	\end{IEEEproof}
\end{lemma}

Note that we have more expectation terms to compute compared to the conventional cellular massive MIMO where a base station only serves the UEs in its cell. In the cellular case, the UEs in a cell usually are assigned orthogonal pilot sequences. Hence, pilot contamination results from the other cells' signals. However, in cell-free massive MIMO, each AP serves all the UEs subsets of which are sharing the same pilot and this in turn results in more terms that appear in both the average harvested energy and SE expressions.

\section{Max-Min Fair Joint LSFD and Power Control}

We want to maximize the minimum SE among the UEs by adjusting both the downlink WPT and uplink WIT power control coefficients and the LSFD weights. 

For the considered optimization, the transmission power of the $l^{\textrm{th}}$ AP during the downlink WPT phase, $P_{l}^E$ in \eqref{eq:Pl} cannot exceed the long-term maximum power limit $\rho_d$ in \eqref{eq:max_power}. Furthermore, we require that the $k^{\textrm{th}}$ UE's total uplink transmission energy, $\tau_u\eta_k+\tau_p\rho_p$ is upper bounded by the harvested energy $E_k$ in \eqref{eq:Ek}. Then, the max-min fairness SE optimization problem is cast as
\begin{align}
& \underset{\left\{\bm{a}_k,\eta_k,p_{kl}\right\},t}{\text{maximize}} \ \ \ t \label{eq:objective} \\
& \text{subject to} \ \ \text{SINR}_k\left(\bm{a}_k,\left\{\eta_i\right\}\right)\geq t, \ \ \ k=1, \ldots,K, \label{eq:constraint1} \\
&\hspace{1.6cm}P_l^E\left(\left\{p_{il}\right\}\right)\leq \rho_d, \ \ \ l=1,\ldots,L, \label{eq:constraint2} \\
&\hspace{1.6cm}\tau_u\eta_k+\tau_p\rho_p\leq E_k\left(\left\{p_{il^{\prime}}\right\}\right), \  \ k=1, \ldots, K, \label{eq:constraint3} \\
&\hspace{1.6cm}p_{kl}\geq 0, \ \  l=1,\ldots,L, \ \ \eta_k\geq 0, \ \   k=1, \ldots, K, \label{eq:constraint4}
\end{align} 
where  $\text{SINR}_k$ is from \eqref{eq:sinr} and $t$ is the SINR that all UEs achieve. Note that this problem is neither convex nor manageable in terms of finding the global optimum solution due to the highly-coupled variables. However, an alternating optimization approach can be developed in an efficient manner where an improved solution is obtained at each step with guaranteed convergence. The motivation for the alternating approach is explained as follows. Note that $P_l^E$ is a linear function of $\{p_{il}\}$. Furthermore, the numerator and denominator of $\text{SINR}_k$ are linear in $\{\eta_i\}$,  given the LSFD vectors $\bm{a}_k$, for $k=1,\ldots,K$. 

The harvested energy for the $k^{\textrm{th}}$ UE, $E_k$ in \eqref{eq:Ek}, is a concave function of the input power $I_k$. As shown in the following part, if we can write $I_k$ in (\ref{eq:Ik-lemma1})-(\ref{eq:Ik-lemma2}) as a linear function of optimization variables, the constraint in \eqref{eq:constraint3} can be expressed in terms of second-order cone constraints that are convex. To do this, let us define the following positive semidefinite matrices that have also rank one.
\begin{align}
& \bm{P}_i\triangleq \begin{bmatrix} \sqrt{p_{i1}} \\ \vdots \\ \sqrt{p_{iL}} \end{bmatrix}\begin{bmatrix} \sqrt{p_{i1}} & \hdots & \sqrt{p_{iL}} \end{bmatrix} \succeq 0,  \ \ \ i=1,\ldots,K.
\end{align}
$I_k$ can be expressed as a linear function of the elements of the matrices $\{\bm{P}_i\}$. The  problem in (\ref{eq:objective})-(\ref{eq:constraint4}) for some given $\bm{a}_k$ and $t$ becomes a feasibility problem. We can include the summation of data transmission powers of UEs in the objective to this problem to obtain good feasible solutions that will improve the next stages of the alternating optimization algorithm. After introducing the new optimization variables $\{e_k\}$, the equivalent version of the problem is 
\begin{align}
& \underset{\left\{\eta_k, \ \bm{P}_k,  \ e_k \right\} }{\text{minimize}} \ \  \sum_{k=1}^K\eta_k   \label{eq:objectivea} \\
& \text{subject to} \ \ \left(1+t\right)\eta_{k}\left|\bm{a}_k^H\bm{b}_k\right|^2-t\sum_{i=1}^K\eta_i\bm{a}_k^H\bm{C}_{ki}\bm{a}_k\nonumber\\&\hspace{2.6cm}-t\bm{a}_k^H\bm{D}_k\bm{a}_k \geq 0, \ \ \ k=1, \ldots,K, \label{eq:constraint1a} \\
&\hspace{0.7cm} \sum_{k=1}^K P_k^{ll}\mathbb{E}\left\{\Vert{\bf w}_{kl}\Vert^2\right\}\leq \rho_d, \ \ \ l=1,\ldots,L, \label{eq:constraint2a}  \\
&\hspace{0.7cm} e_k\left(B_kI_k\left(\left\{P_i^{ll^{\prime}}\right\}\right)+C_k\right)\geq 1, \ \ \ k=1, \ldots,K,  \label{eq:constraint3a} \\
&\hspace{0.7cm}\tau_u\eta_k+\tau_p\rho_p\leq \frac{\tau_dA_k}{B_k}-\frac{\tau_dA_kC_k}{B_k}e_k, \ \ \ k=1, \ldots, K, \label{eq:constraint4a} \\
&\hspace{0.7cm}P_k^{ll^{\prime}}\geq 0, \ \  l=1,\ldots,L, \ \ l^{\prime}=1,\ldots,L, \nonumber\\& \hspace{0.7cm}   \eta_k\geq 0, \ \   k=1, \ldots, K, \label{eq:constraint5a} \\
&\hspace{0.7cm} \bm{P}_k\succeq 0, \ \ \ \text{rank}\left(\bm{P}_k\right)=1, \ \ k=1,\ldots,K, \label{eq:constraint6a}
\end{align} 
where $P_i^{ll^{\prime}}$ denote the $\left(l,l^{\prime}\right)$th element of the matrix $\bm{P}_i$. Note that since $B_kI_k\left(\left\{P_i^{ll^{\prime}}\right\}\right)+C_k$ is a positive affine function of the optimization variables $\left\{P_i^{ll^{\prime}}\right\}$, the constraints in \eqref{eq:constraint3a} can be written as second-order cone constraints. Note that  \eqref{eq:constraint4a} is now linear with respect to the optimization variables and the two optimization problems (\ref{eq:objective})-(\ref{eq:constraint4})  and (\ref{eq:objectivea})-(\ref{eq:constraint6a}) are equivalent in the sense that they have the same global optimum solution. To show this, suppose at least one of the inequalities  \eqref{eq:constraint3a} is not satisfied with equality for the optimum solution of (\ref{eq:objectivea})-(\ref{eq:constraint6a}). In this case, the corresponding $e_k>0$ can be reduced until the constraint becomes an equality without affecting the feasibility and the optimality of the solution. This modified solution is also a global optimum for (\ref{eq:objectivea})-(\ref{eq:constraint6a}), and hence for (\ref{eq:objective})-(\ref{eq:constraint4}) since $e_k=1\big/\left(B_kI_k\left(\left\{P_i^{ll^{\prime}}\right\}\right)+C_k\right)$.

We note that the only constraints that destroy the convexity in (\ref{eq:objectivea})-(\ref{eq:constraint6a}) are the rank one constraints in \eqref{eq:constraint6a}. The following lemma shows that we can obtain an optimum solution to  (\ref{eq:objectivea})-(\ref{eq:constraint6a})  by solving it without those constraints.

\begin{lemma}\label{rank} Let $\left\{\bm{P}_k^{\star}\right\}$ denote the optimum matrices for the problem (\ref{eq:objectivea})-(\ref{eq:constraint6a}) with dropped rank one constraints. Then, the rank-one matrices $\left\{\bm{P}_k^{\star\star}\right\}$ defined as
	 \begin{align}
	& \bm{P}_k^{\star\star}\triangleq \begin{bmatrix} \sqrt{{P_k^{11}}^\star} \\ \vdots \\ \sqrt{{P_k^{LL}}^\star} \end{bmatrix}\begin{bmatrix} \sqrt{{P_k^{11}}^\star} & \hdots & \sqrt{{P_k^{LL}}^\star} \end{bmatrix},   k=1,\ldots,K, \label{eq:Pk}
	\end{align} 
	constitute another optimum solution to the original problem (\ref{eq:objectivea})-(\ref{eq:constraint6a}). 
	\begin{IEEEproof} Please see the Appendix~\ref{lemma4proof}.
		\end{IEEEproof}
	\end{lemma}

Using Lemma~4, we can obtain the global optimum solution of  the problem (\ref{eq:objectivea})-(\ref{eq:constraint6a}) by removing the non-convex rank constraints and solving it with convex programming. Hence, for a given set of the LSFD vectors, the original problem in (\ref{eq:objective})-(\ref{eq:constraint4}) can be shown to be quasi-convex and its global optimum solution can be found using bisection search over $t$ by solving a series of convex programming problems \cite{cell_free_vs_small_cell}. Furthermore, the LSFD vector $\bm{a}_k$ only affects the SINR of the $k^{\textrm{th}}$ UE and can be found in closed form for the given uplink power coefficients $\{\eta_i\}$ by maximizing a generalized Rayleigh quotient \cite{ozge_cell_free}. Using these observations, we propose the alternating optimization algorithm which combines the closed-form LSFD vectors with the bisection search over minimum SINR as stated in Algorithm~1.

\begin{figure}[t!]
	\begin{center}
		\hspace{-1cm}
		\begin{overpic}[width=7.6cm,tics=10]{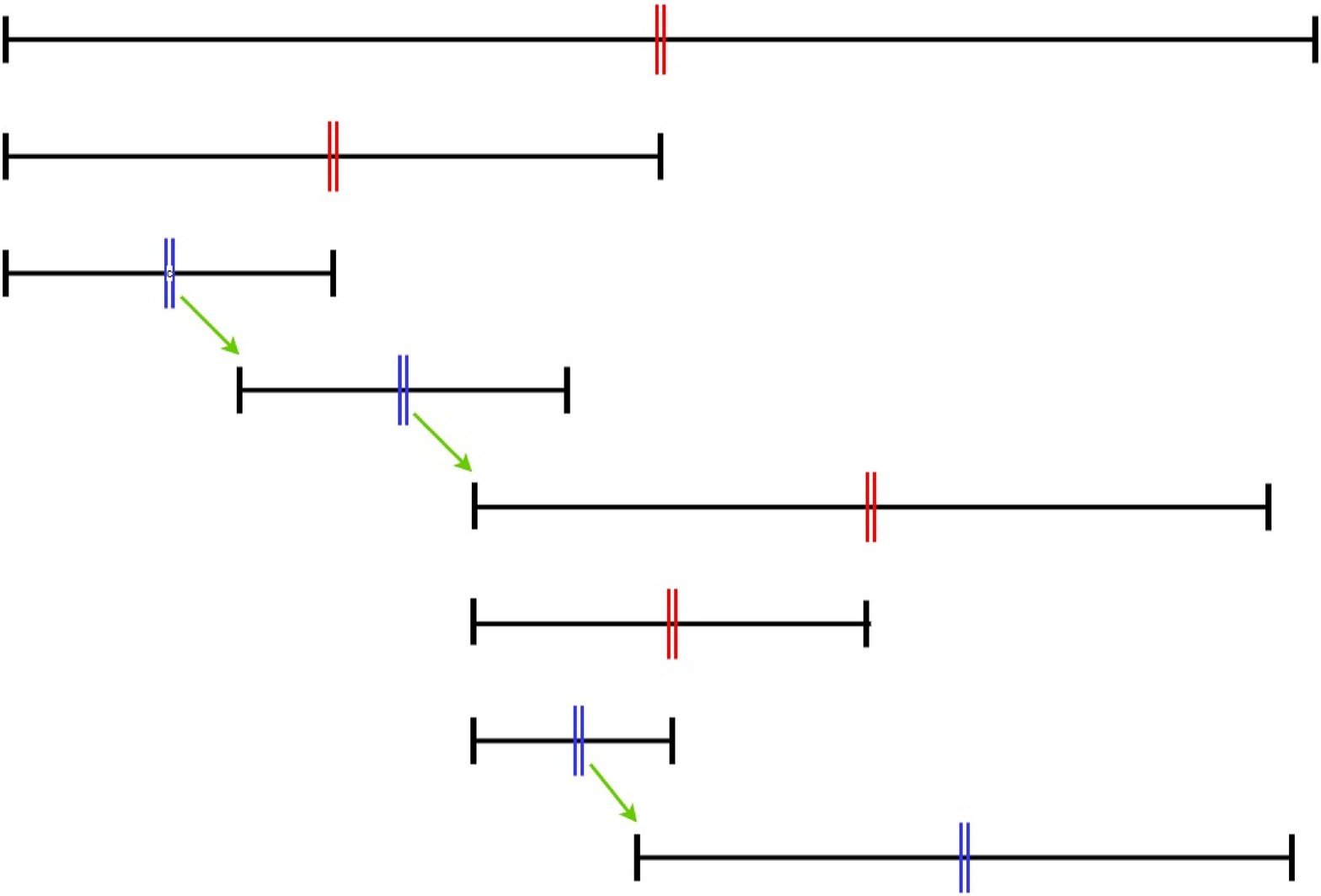}
				\put(-2,59.3){\scriptsize $t_{\text{min}}^{(0)}$}
			 \put(98,59.3){\scriptsize $t_{\text{max}}^{(0)}$}
			 			 \put(48,59.3){\scriptsize $t^{(1)}$}
			 	\put(-2,50.3){\scriptsize $t_{\text{min}}^{(1)}$}
			 				 	\put(48,50.3){\scriptsize $t_{\text{max}}^{(1)}$}
	\put(23,50.3){\scriptsize $t^{(2)}$}
		\put(-2,41.2){\scriptsize $t_{\text{min}}^{(2)}$}
	\put(23,41.2){\scriptsize $t_{\text{max}}^{(2)}$}
	\put(10.5,41.2){\scriptsize $t^{(3)}$}
		\put(12,32.2){\scriptsize $t_{\text{min}}^{(3)}=t^{\star}$}
	\put(41,32.2){\scriptsize $t_{\text{max}}^{(3)}=\lambda t^{\star}$}
	\put(29,32.2){\scriptsize $t^{(4)}$}
		\put(34,24){\scriptsize $t_{\text{min}}^{(4)}=t^{\star}$}
	\put(94,24){\scriptsize $t_{\text{max}}^{(4)}=\lambda t^{\star}$}
	\put(65,24){\scriptsize $t^{(5)}$}
		\put(34,15){\scriptsize $t_{\text{min}}^{(5)}$}
	\put(64,15){\scriptsize $t_{\text{max}}^{(5)}$}
	\put(50,15){\scriptsize $t^{(6)}$}
		\put(34,5.7){\scriptsize $t_{\text{min}}^{(6)}$}
	\put(49.5,5.7){\scriptsize $t_{\text{max}}^{(6)}$}
	\put(42,5.7){\scriptsize $t^{(7)}$}
		\put(46,-3){\scriptsize $t_{\text{min}}^{(7)}=t^{\star}$}
	\put(96,-3){\scriptsize $t_{\text{max}}^{(7)}=\lambda t^{\star}$}
	\put(72,-3){\scriptsize $t^{(8)}$}
			\end{overpic}
		\end{center}
		\caption{An example of the search procedure of the  modified bisection search in Algorithm~1. The leftmost and the rightmost black lines denote $t_{\text{min}}^{(r-1)}$ and $t_{\text{max}}^{(r-1)}$, respectively, at the $r^{\textrm{th}}$ iteration. The double lines show $t^{(r)}$ and its color is blue if the problem in Step 4 of Algorithm 1 is feasible and red otherwise. The green arrows denote the possible increase in the minimum SINR after applying LSFD.  
		} \label{fig:bisection}
\end{figure}

Algorithm~1 is a modified bisection search algorithm over the minimum SINR. The reason we call it modified will be explained as follows. Fig.~\ref{fig:bisection} shows an example search procedure of Algorithm~1 for the first 8 iterations. The algorithm starts by setting the lower and upper bound for the minimum SINR, i.e., $t_{\text{min}}=0$ and $t_{\text{max}}$. The initial value of $t_{\text{max}}$ in Algorithm 1 can be taken as an upper bound on $t$ for the problem (\ref{eq:objective})-(\ref{eq:constraint4}). A simple upper bound can be obtained by supposing there is only one UE in the setup and maximizing the SINR of that UE. If we focus on the $k^{\textrm{th}}$ UE, the harvested energy, $E_k$ in \eqref{eq:Ek} is maximized by setting $p_{kl}=\rho_d/\mathbb{E}\left\{\Vert{\bf w}_{kl}\Vert^2\right\}$ and $p_{il}=0$, $\forall i\neq k$  by (\ref{eq:max_power})-(\ref{eq:Pl}). Let $E_k^{\star}$ denote the value of harvested energy for this setting. To maximize the $\text{SINR}_k$, we equate the total uplink transmission energy for the $k^{\textrm{th}}$ UE to the harvested energy $E_k^{\star}$ in \eqref{eq:constraint3} and obtain the data power control coefficient as $\eta_k^{\star}$. We set all other uplink power control coefficients to zero, i.e., $\eta_i=0$, $\forall i \neq k$. After maximizing the obtained generalized Rayleigh quotient for the $k^{\textrm{th}}$ UE, we obtain $\text{SINR}_k^{\star}$. If we repeat this procedure for each UE, we can obtain a proper upper bound for the initialization of Algorithm 1 as follows:
\begin{align}
t_{\text{max}}=\min_{k}\text{SINR}_k^{\star}. \label{eq:tmax}
\end{align}
Then, the convex problem in (\ref{eq:objectivea})-(\ref{eq:constraint6a}) without rank constraints is solved for the fixed value of the LSFD vectors $\{\bm{a}_k\}$ and minimum SINR $t$. If the problem is not feasible, which is denoted by red double lines in Fig.~\ref{fig:bisection}, the conventional bisection search procedure is applied. However, if the problem is feasible, which is denoted by blue double lines, then Steps 6-9 in Algorithm~1 are applied consecutively. The motivation for the power scaling in  Step 7 of the Algorithm 1 is to increase the minimum SINR that every UE attains. Since the problem (\ref{eq:objectivea})-(\ref{eq:constraint6a}) does not take into account the max-min SINR (constant $t$), the optimum solution of it may be scaled for a potential increase in $t$, which is useful in the next iterations of the Algorithm 1.

Note that in Step 9 of Algorithm 1, we change $t_{\text{max}}$ to $\lambda t^{\star}$ (Here, $\lambda>1$ is a scaling parameter to extend the bisection interval). The reason for this update is that after LSFD, it may be possible to obtain feasible solution with $t$ larger than the $t_{\text{max}}$ that is set at the previous infeasible iterations. This potential increase in the minimum SINR is shown with a green arrow in Fig.~\ref{fig:bisection}. Due to this dynamic interval adaptation in the bisection search owing to LSFD, we call our algorithm a modified bisection search. We note that the objective function of the problem (\ref{eq:objective})-(\ref{eq:constraint4}) is upper bounded as shown above and an improved solution is obtained at each iteration. Hence, Algorithm 1 converges.

We note that the problem in (\ref{eq:objectivea})-(\ref{eq:constraint6a}) can be expressed in standard semidefinite programming form by expressing the second-order cone constraints in terms of positive semidefinite matrices and arranging the inequalities. Solving this problem dominates the computational complexity of Algorithm~1. A semidefinite program can be solved by using polynomial time algorithms and the interior point techniques are among the most widely used methods. It has a worst-case complexity of $\tilde{\mathcal{O}}\left(m\left(m^2+n^{\omega}+mns\right)\log\left(1/\delta\right)\right)$ where $n$, $m$, and $s$ are the size of the positive semidefinite matrices, the number of inequality constraints, and the maximum number of non-zero entries in each row of the weighting matrices, respectively, in the standard formulation \cite{sdp_complexity}. The parameter $\delta$ is the solution accuracy and $\omega$ is the exponent of matrix multiplication. Note that $\tilde{\mathcal{O}}$ notation corresponds to the big-O notation while excluding the logarithmic factors. After arranging the problem in (\ref{eq:objectivea})-(\ref{eq:constraint6a}) in a standard semidefinite programming form, we have $n=K(L+3)$, $m=K\left(L^2-L\right)/2+L+4K$, and $s=KL$, that may lead to huge complexity as the number of APs and UEs in the network increases. The computational complexity of semidefinite programming can be reduced by the newly proposed methods \cite{sdp_complexity} and by exploiting the highly sparse structure of the most constraints such as simple bound constraints. However, this is outside the scope of this paper. The main purpose of this paper is to demonstrate the achievable performance in terms of max-min fairness with several channel estimation, transmission and reception strategies. We emphasize that the obtained closed-form results can be exploited for simpler power control schemes as well.

\begin{algorithm}
	\caption{ Modified Bisection Search for Max-Min Fair LSFD and Power Control}
	\begin{algorithmic}[1]
		\State {\bf Initialization:} Set $t_{\text{min}}=0$ and $t_{\text{max}}$  as in \eqref{eq:tmax}, respectively. Initialize $\bm{a}_k$ as all ones vector for $k=1,\ldots,K$. 
		\While{$t_{\text{max}}-t_{\text{min}}>\epsilon$} \Comment{$\epsilon>0$ determines the solution accuracy.}
		\State Set $t=\frac{t_{\text{min}}+t_{\text{max}}}{2}$.
		\State Solve the convex problem in (\ref{eq:objectivea})-(\ref{eq:constraint6a}) without rank constraints and by taking $\{\bm{a}_k\}$ and $t$ as constant.
		\If{feasible}
		\State $\bullet$ Set the power control coefficients as the solution of this problem. \Comment{ the diagonal elements of the matrices $\left\{\bm{P}_k\right\}$.}
		\State $\bullet$ Scale all the power control coefficients $\{p_{kl}\}$ and $\{\eta_k\}$ so that at least one of the constraints in \eqref{eq:constraint2} and \eqref{eq:constraint3}, respectively, are satisfied with equality. 
		\State $\bullet$ Obtain the optimum $\{\bm{a}_k\}$ by maximizing each UE's SINR as a generalized Rayleigh quotient.
		\State   $\bullet$ Set $t_{\text{min}}=t^{\star}$ and $t_{\text{max}}=\lambda t^{\star}$ where $t^{\star}$ is the minimum of the SINRs after applying LSFD. 
		\Else
		\State  Set  $t_{\text{max}}=t$.
		\EndIf
		\EndWhile
		\State {\bf Output:} Downlink power coefficients  $\left\{p_{kl}\right\}$, uplink power coefficients $\left\{\eta_k\right\}$, LSFD vectors $\left\{\bm{a}_k\right\}$, minimum SINR $t$.
	\end{algorithmic}
\end{algorithm}

{\bf Remark:} We note that the steps of Algorithm 1 are intended for coherent energy transmission and the non-linear energy harvesting model in \eqref{eq:Ek}. For the linear energy harvesting model, there is no need for introducing the optimization variables $\left\{e_k\right\}$. Similarly, the non-diagonal elements of the matrices $\bm{P}_k$ are not used for non-coherent energy harvesting model since the input power to the harvester is simply a linear function of $\left\{p_{kl}\right\}$. Hence, for other scenarios, the optimization problem in  (\ref{eq:objectivea})-(\ref{eq:constraint6a}) can be simplified accordingly.

\section{Numerical Results}

In this section, we will quantify the SE for different energy harvesting models and transmission schemes together with various setups.  The 3GPP indoor hotspot (InH) model in \cite{3gpp} is used with a 3.4 GHz carrier frequency and 20\,MHz bandwidth. The large-scale fading coefficients, shadowing parameters, probability of LOS, and the Rician factors are simulated based on \cite[Table B.1.2.1-1, B.1.2.1-2, B.1.2.2.1-4]{3gpp}. The APs are uniformly distributed in a 100\,m$\times$100\,m square.\footnote{The correlations between shadowing, terminal positions and Rician factors in \cite{3gpp} are neglected for simplicity.} For each setup, the UEs are randomly dropped and a 4\,m height difference between APs and UEs is taken into account when calculating distances. The noise variance is $\sigma^2=-96$ dBm. The uplink pilot transmission power is $-40$\,dBm. The total number of samples per coherence interval is $\tau_c=200$ with $\tau_p=5$, $\tau_d=25$, and $\tau_u=170$ unless otherwise stated. The constant part of the LOS components are generated by assuming a uniform linear array in the far field of the users with half wavelength antenna spacing. For each scenario, 500 random setups corresponding to different user locations are considered where the number of UEs is $K=20$ unless otherwise stated.

\begin{figure}[t!]
		\begin{center}
			\includegraphics[trim={1.6cm 0.1cm 3cm 1cm},clip,width=8.8cm]{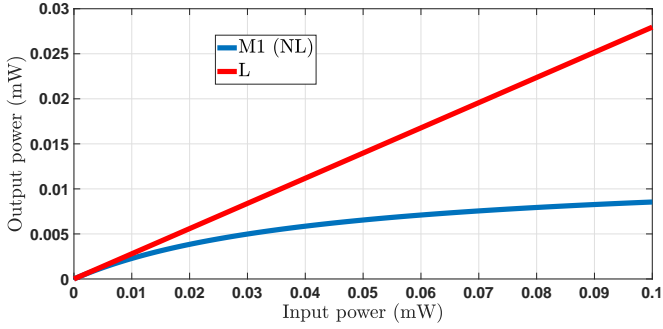}
			\caption{The non-linear energy harvesting model (M1) from \cite{model1} and the corresponding linear energy harvesting model (L).} \label{fig:model1}
		\end{center}
	\end{figure}

\begin{figure}[t!]
		\begin{center}
			\includegraphics[trim={1.6cm 0.1cm 3cm 1cm},clip,width=8.8cm]{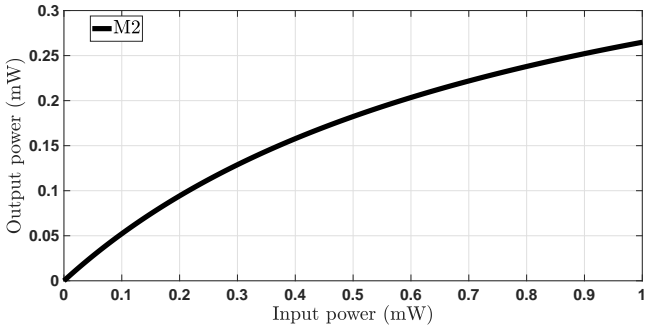}
			\caption{The non-linear energy harvesting model (M2) from \cite{model2}.} \label{fig:model2}
		\end{center}
\end{figure}

In the simulations, we consider both coherent and non-coherent energy transmission schemes, which are labeled as C and NC in the figures, respectively. We consider the linear and non-linear energy harvesting models in \cite{revise1}. The parameters for the non-linear energy harvesting model in \eqref{eq:Ek} are set to be the same for each UE. Two different non-linear models are considered to describe the saturation effect of a practical energy harvesting circuit. The parameters for the first model, which is labeled as M1 in the figures, are given as $A_k=10^3(ac-b)$, $B_k=10^6c$, and $C_k=10^3c^2$ where $a=0.3929$, $b=0.01675$, and $c=0.04401$ are obtained in \cite{revise1} by fitting to measurement data in \cite{model1}. As shown in Fig.~\ref{fig:model1}, the harvested power starts to saturate at around 0.1\,mW. For most of the simulations, we consider this model. To obtain a linear energy harvesting model as a benchmark to M1, we set $B_k$ to zero and this selection always results in more harvested power as shown in Fig.~\ref{fig:model1}. It is labeled as L in the remaining figures. The second non-linear energy harvesting model parameters are $a=2.463$, $b=1.635$, and $c=0.826$, which are obtained by fitting to measurement data in \cite{model2}. This model is labeled as M2 and the harvested power saturates at approximately 10 times higher input power compared to M1, as shown in Fig.~\ref{fig:model2}. Hence, M2 behaves approximately as a linear energy harvesting model around 0.1\,mW where the harvested power for M1 shows a high non-linear distortion. In the last part of the simulations, we will compare the SE obtained by M1 and M2.   

In the first scenario, we consider $L=36$ APs, each with $N=8$ antennas. The maximum power of each AP is $\rho_d=10/36$\,W corresponding to 10\,W of total maximum power for the considered cell-free network. MMF stands for the proposed max-min fairness optimization. We compare the proposed max-min fairness optimization with a simpler power control that is inspired by the fractional power control (FPC) scheme for the downlink information transmission in \cite{giovanni}. For this scheme, the power control coefficient $p_{kl}$ is proportional to $1\mathbin{/}\sqrt{\mathbb{E}\left\{\left\vert{\bf w}_{kl}\right\vert^2\right\}}$ and they are scaled such that the total transmission power is $\rho_d$ for each AP in accordance with the power control scheme \cite{giovanni}. Each UE's power control coefficient $\eta_k$ is adjusted such that the total uplink transmission energy is equal to the harvested energy in the downlink. The MR precoding and decoding vectors are obtained by the LMMSE-based channel estimation scheme. 

\begin{figure}[t!]
		\begin{center}
			\includegraphics[trim={2.1cm 0.1cm 3cm 1cm},clip,width=8.8cm]{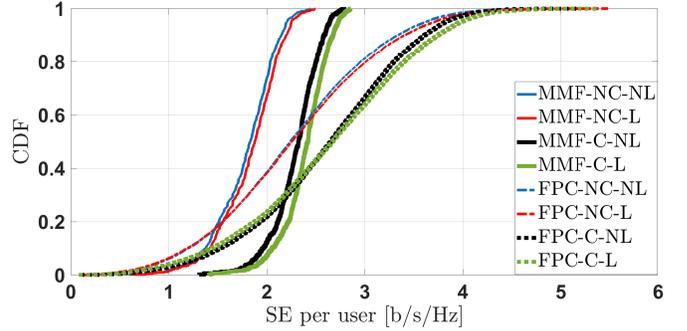}
			\caption{The CDF of SE per user for the proposed MMF and FPC in \cite{giovanni}, $L=36$, $N=8$, and LMMSE-based channel estimation.} \label{fig:sim1}
		\end{center}
	\end{figure}

In Fig.~\ref{fig:sim1}, we plot the cumulative distribution function (CDF) of the individual SE per UE. We notice that the 90\% and 95\% likely SE (i.e., where the CDF is 0.1 and 0.05, respectively) for coherent energy transmission is 42\% and 76\% higher for the proposed MMF design  in comparison to FPC for linear energy harvesting model. The 90\% and 95\% likely SE improvement over FPC is around 28\% and 54\%, respectively for non-linear energy harvesting. The higher improvement with the linear energy harvesting model can be explained as follows. Since more energy can be harvested with the linear model, the MMF optimization results in higher minimum SE among the users. However, for FPC, the UEs with good channel conditions (corresponding to the upper tail of the CDF curves) are able to obtain higher SE compared to non-linear energy harvesting since these users are expected to operate the saturation region of M1 in Fig.~\ref{fig:model1} where the gap between M1 and L is higher compared to the left region of the graph. Hence, with FPC that does not consider max-min fairness, the UEs with high signal-to-noise ratio (SNR) benefit from the linear energy harvesting whereas the low-SNR UEs attain less SE in order to compensate. On the other hand, coherent energy transmission results in significantly higher SE compared to the non-coherent one due to the reasoning explained in Section~\rom{4}.

\begin{figure}[t!]
		\begin{center}
			\includegraphics[trim={2.1cm 0.1cm 3cm 1cm},clip,width=8.8cm]{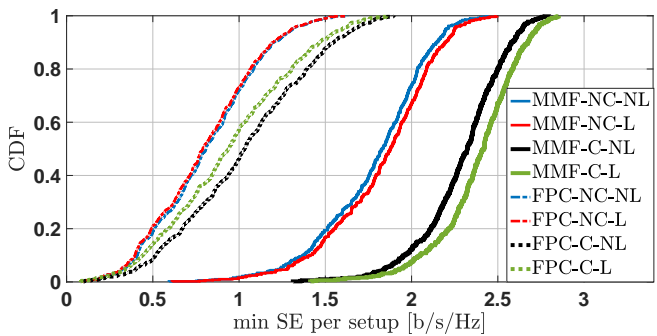}
			\caption{The CDF of minimum SE per setup for the same scenario as Fig.~\ref{fig:sim1}.} \label{fig:sim2}
		\end{center}
\end{figure}

In order to see the fairness improvement of the proposed algorithm, we plot the CDF of the minimum SE of all the UEs per setup in Fig.~\ref{fig:sim2} for the same scenario. The minimum SE of the network improves substantially and larger SE is guaranteed for all the UEs. In the following experiments, we quantify the impact of several parameters on the SE for the proposed MMF design.

\begin{figure}[t!]
		\begin{center}
			\includegraphics[trim={2.1cm 0.1cm 3cm 1cm},clip,width=8.8cm]{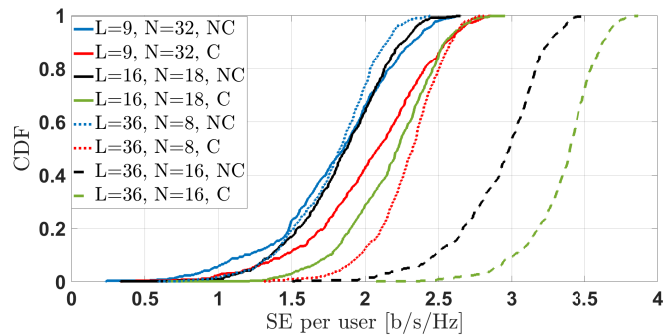}
			\caption{The CDF of SE per user for LMMSE-based channel estimation.} \label{fig:sim3}
		\end{center}
	\end{figure}

In Fig.~\ref{fig:sim3}, we quantify the impact of the number of APs, $L$, and antennas per AP, $N$ for the first non-linear energy harvesting model in Fig.~\ref{fig:model1}, M1. The maximum transmission power for each AP is $\rho_d=10\mathbin{/}L$\,W. Hence, maximum total transmit power for the whole AP network is $10$\,W for a fair comparison. For the first six lines in Fig.~\ref{fig:sim3}, the number of total antennas throughout all the area is $LN=288$. We notice that the 90\% likely SE is improved by 35\% and 16\% for coherent and non-coherent energy transmission, respectively, by increasing the number of APs from $L=9$ to $L=36$. However, there is a slight performance decrease in some regions of the CDF curve when we increase the number of APs by keeping the total number of antennas the same especially for NC case. That the improvement is not visible as in increasing $L=9$ to $L=16$ is most probably due to the adverse effect of the increased local power constraints in \eqref{eq:constraint2}. However, if we increase the number of antennas per AP to $N=16$ for $L=36$, we now see the positive impact of jointly increasing the number of APs and total number of antennas, $LN$, where each UE's SE is significantly improved. Another important observation from Fig.~\ref{fig:sim3} is that the SE gap between coherent and non-coherent energy transmission increases with the number of APs since coherent combining of energy signals transmitted from different APs  supplies more power to be harvested at each UE under the same total power constraint.

\begin{figure}[t!]
		\begin{center}
			\includegraphics[trim={2.1cm 0.1cm 3cm 1cm},clip,width=8.8cm]{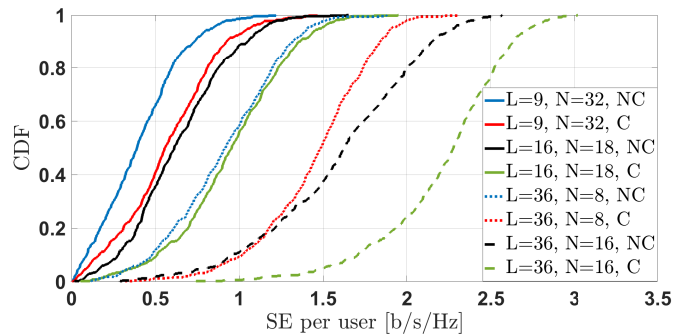}
			\caption{The CDF of SE per user for LS-based channel estimation.} \label{fig:sim4}
		\end{center}
\end{figure}

In Fig.~\ref{fig:sim4}, we repeat the previous simulation with LS-based channel estimation. Since LS-based estimation utilizes no channel statistics on the contrary to the LMMSE-based estimation, significantly less SE is achieved compared to Fig.~\ref{fig:sim3}. However, it is possible now to improve the SE significantly  by increasing the number of APs to $L=36$ by keeping the total number of antennas in the network fixed at $LN=288$. The 90\% likely SE is 6 and 5 times greater compared to the case of $L=9$ APs for coherent and non-coherent energy transmission, respectively. In contrast to the LMMSE-based channel estimation, increasing the number of APs results in significantly higher SE for all the UEs. This may be due to that placing APs in a denser manner improves the channel estimation quality of LS and the SE is continued to improve although the number of local power constraints has been increased. 

\begin{figure}[t!]
		\begin{center}
			\includegraphics[trim={2.1cm 0.1cm 3cm 1cm},clip,width=8.8cm]{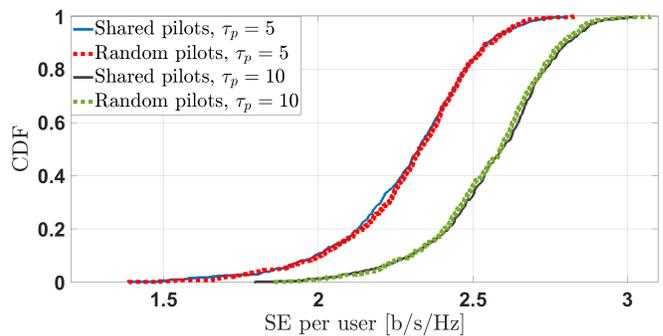}
			\caption{The CDF of SE per user for shared and random pilot sequences with different lengths.}
			\label{fig:sim5a}
		\end{center}
	\end{figure}

Throughout the following experiments, we consider the setup with $L=36$ and $N=8$. In Fig.~\ref{fig:sim5a}, we compare the SE obtained when the UEs are either sharing a set of mutually orthogonal pilots or have randomly generated partially overlapping pilots. The former case is the one we have used in the analytical part of this paper and it coincides with most previous works in the cell-free massive MIMO literature. In case of random pilot sequences \cite{wpt3}, each UE's pilot sequence is randomly generated from i.i.d. standard Gaussian distribution and is normalized such that its squared norm is $\tau_p$. Unlike the i.i.d. Rayleigh fading channels that enable estimation of the channel gain of each AP antenna in an independent manner \cite{wpt3}, there is a common random phase shift that affects each AP antenna jointly in the more general channel model in (1). The randomly generated pilot sequences require in this case construction of $N\tau_p\times N\tau_p$ statistical matrices and inversion of them for the LMMSE channel estimator. Hence, the closed-form expressions will be different than the ones we have derived in this paper. To quantify the difference between the shared mutually orthogonal and random pilot sequences, we use Monte Carlo estimation for the SE and average harvested energy expressions.
	
	In Fig.~\ref{fig:sim5a}, we consider the non-linear energy harvesting model M1 with coherent energy transmission and LMMSE-based channel estimation. As the figure shows, there is a negligible performance difference between the two pilot signaling schemes for the two considered  pilot lengths: $\tau_p=5$ and $\tau_p=10$. We note that the pilot contamination is not totally undesirable thing in wireless-powered cell-free networks since it also contributes to the harvested energy as can be seen from \eqref{eq:Ik-lemma1}-\eqref{eq:Ik-lemma2}  and thus there is an implicit trade-off. As a result, we do not observe a substantial performance gap that may result from the diversity of the random pilot scheme. However, the shared pilot scheme has lower computational complexity since we can apply the pilot de-spreading in \eqref{eq:suff-stats} and this makes it preferable compared to the case of random pilots.

\begin{figure}[t!]
		\begin{center}
			\includegraphics[trim={2.1cm 0.1cm 3cm 1cm},clip,width=8.8cm]{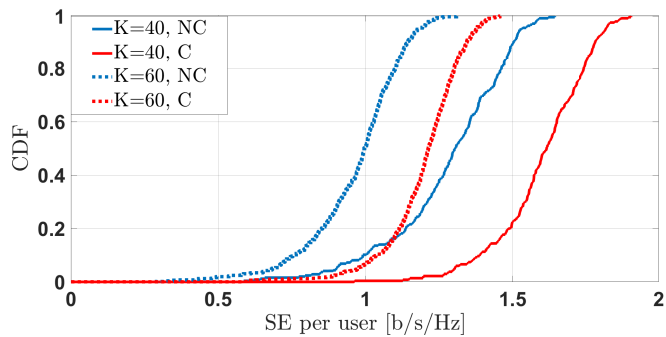}
			\caption{The CDF of SE per user for different number of UEs.}
			 \label{fig:sim6a}
		\end{center}
\end{figure}

	 In Fig.~\ref{fig:sim6a}, we evaluate the SE per user for the scenario with $\tau_p=5$ and different number of UEs, namely $K=40$ and $K=60$. We consider the non-linear energy harvesting model M1 and LMMSE-based channel estimation as before. As expected, the SE that can be provided to the worst UE in the network decreases with the increase in the number of UEs. In very rare scenarios, the SE is almost zero as can be observed from the lower tail of CDF curves. This is because there is a potential risk that the most unfortunate UE in the coverage area may have very bad channel conditions due to shadowing and the other UEs are forced to reduce their power to maximize its SE.  However, it is possible to attain a practically reasonable SE for most of the cases and UEs. In the least favorable scenario with NC and $K=60$ UEs, the median SE is 1\,b/s/Hz and this value can be increased further by coherent energy transmission and deploying more antennas in the network. In fact, $L=36$ APs with $N=8$ antennas each (i.e., $LN=288$ antennas in total) can provide satisfactorily high SE to $K=60$ UEs, which is a relatively high number of UEs for conventional massive MIMO setups.

\begin{figure}[t!]
		\begin{center}
			\includegraphics[trim={2.1cm 0.1cm 3cm 1cm},clip,width=8.8cm]{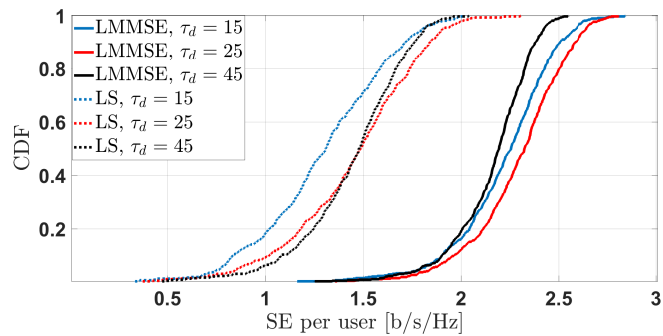}
			\caption{The CDF of SE per user for non-linear energy harvesting model M1 and coherent energy transmission.} \label{fig:sim5}
		\end{center}
	\end{figure}

 In Fig~\ref{fig:sim5}, we plot the CDF curves of the SE per user for different number of downlink energy symbols, $\tau_d$, by keeping $\tau_p=5$ and $\tau_c=200$ constant as before. The number of uplink symbols changes with $\tau_d$ as $\tau_u=200-5-\tau_d$. We consider only the non-linear energy harvesting model M1 and the coherent energy transmission. As can be seen from Fig.~\ref{fig:sim5}, there is not a large performance difference for $\tau_d=15$, $\tau_d=25$, and $\tau_d=45$ for both channel estimation methods. This is due to the trade-off between harvested energy duration and duration of uplink information transmission. As $\tau_d$ increases, the harvested energy increases linearly with $\tau_d$, however, $\tau_u$ decreases by reducing the pre-log factor in the SE formula. Hence, an increase in harvested energy allows each UE to transmit with a higher uplink power but with a reduced uplink information duration.

\begin{figure}[t!]
		\begin{center}
			\includegraphics[trim={2.1cm 0.1cm 3cm 1cm},clip,width=8.8cm]{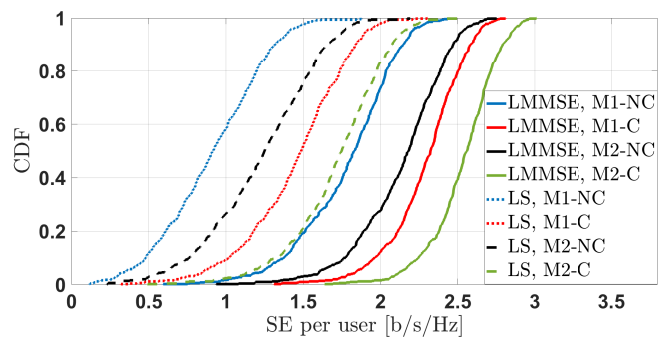}
			\caption{The CDF of SE per user for different non-linear energy harvesting models.} \label{fig:sim6}
		\end{center}
\end{figure}

In Fig.~\ref{fig:sim6}, we compare the SE performance of the two non-linear energy harvesting models, i.e., M1 and M2, which are presented in Figs.~\ref{fig:model1} and \ref{fig:model2}, respectively. For both channel estimation methods and energy transmission schemes, M2 that has a higher input power range before saturation effect, provides a higher SE compared to M1. In the 0-0.1\,mW input power region in Fig.~\ref{fig:model1}, M2 almost behaves as a linear energy harvester with a greater harvested power, hence the improvement in SE is not surprising. This shows us the importance of selecting the right energy harvesting circuit parameters when analyzing these systems.


\begin{figure}[t!]
		\begin{center}
			\includegraphics[trim={2.1cm 0.1cm 3cm 1cm},clip,width=8.8cm]{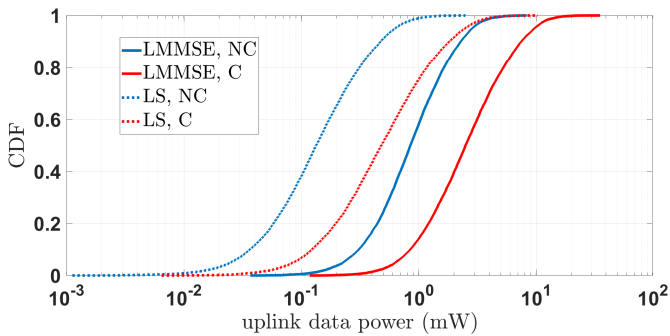}
			\caption{The CDF of uplink data power for $L=36$ and $N=8$ with non-linear energy harvesting model M1.} \label{fig:sim7}
		\end{center}
	\end{figure}

\begin{figure}[t!]
		\begin{center}
			\includegraphics[trim={1.6cm 0cm 3cm 0.1cm},clip,width=8.8cm]{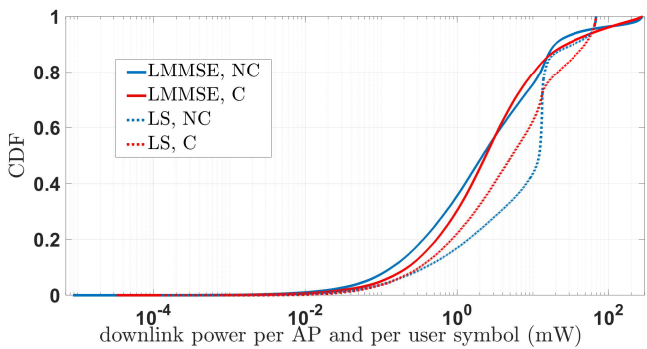}
			\caption{The CDF of downlink power for $L=36$ and $N=8$ with non-linear energy harvesting model M1.} \label{fig:sim8}
		\end{center}
\end{figure}

As a final simulation, we plot the CDF of uplink and downlink data transmission powers for the non-linear energy harvesting model M1 in Fig.~\ref{fig:sim7} and Fig.~\ref{fig:sim8}, respectively. Note that the maximum uplink transmission power for each UE is determined by the corresponding harvested power. As can be seen from Fig.~\ref{fig:sim7}, more uplink power is utilized to increase the minimum SE of the network for C transmission compared to NC one since the maximum available power for the UEs are expected to increase with C downlink transmission. In addition, having a better channel estimate (LMMSE compared to LS) increases the uplink data power consistently. When it comes to downlink power distribution, Fig.~\ref{fig:sim8} plots the per AP and per UE symbol downlink transmission powers. For $L=36$ APs, the maximum available power for each AP is around 278\,mW and for LMMSE-based channel estimation, there are some scenarios where some APs use its almost full power for only one UE. However, this case does not happen for LS-based channel estimation, probably to poorer channel estimation quality. 

\section{Conclusion}

In this paper, the uplink SE of the wireless-powered cell-free massive MIMO has been derived for MR processing with LMMSE and LS-based channel estimations and LSFD at the CPU. The channels were assumed to follow a practical Rician fading distribution with unknown phase-shifted LOS components in each coherence block. The UEs harvest energy from the downlink RF signals and use a portion of it for the uplink data transmission. For a non-linear energy harvesting model, whose parameters can be fitted to different real data measurements with saturation effects, the average harvested energy has been derived for the LMMSE and LS-based channel estimations and two different energy transmission schemes: coherent and non-coherent. Using the derived uplink SE and harvested energy in the downlink, we optimized both the downlink WPT and uplink WIT power control coefficients together with the LSFD weights to maximize the minimum guaranteed SE for all the UEs. An alternating optimization algorithm is proposed for solving the non-convex problem. To solve the resulting non-convex sub-problems efficiently, the problem has been transformed into a new form with additional variables and constraints. 

Several simulations were carried out with practical non-linear and conventional linear energy harvesting models. The results show that the proposed MMF algorithm significantly improves the fairness by providing greater SE to the weakest UEs compared to another state-of-the-art power control scheme that was originally proposed for downlink information transmission. Furthermore, coherent energy transmission increases the SE of each UE in comparison to its non-coherent counterpart with an additional burden of downlink synchronization among the APs. The performance improvement becomes more visible with the increase in the number of APs that further improves the SE. We also note that the effect of the downlink energy symbol length and the pilot signaling scheme on the SE is not as significant as the other system parameters.

\appendices
\section{Useful Lemmas}
\begin{lemma}\label{applemma1}
	\cite[Lemma 2]{emil_nonideal}. Consider the  random vector ${\bf u}\in \mathbb{C}^N$ that is distributed as ${\bf u} \sim \mathcal{N}_{\mathbb{C}}\left({\bf 0}_N,{\bf A}\right)$. For a deterministic matrix ${\bf B}\in \mathbb{C}^{N\times N}$, it holds that
	\begin{align}
	&\mathbb{E}\left\{\left|{\bf u}^H{\bf B}{\bf u}\right|^2\right\}= \left|\tr\left({\bf A}{\bf B}\right)\right|^2+\tr\left({\bf A}{\bf B}{\bf A}{\bf B}^H\right).
	\end{align}
\end{lemma}

\begin{lemma}\label{applemma2} Consider the vectors ${\bf x}=e^{j\theta_x}{\bf \bar{x}}+\sigma_x{\bf w} \in \mathbb{C}^{N}$ and ${\bf y}=\alpha{\bf x}+{\bf z} \in \mathbb{C}^{N}$, where ${\bf \bar{x}} \in \mathbb{C}^{N}$ is deterministic and $\theta_x$ is uniformly distributed in the interval $[0,2\pi)$. $\sigma_x\geq 0$ and $\alpha$ are some real deterministic scalars and ${\bf w} \sim \mathcal{N}_{\mathbb{C}}({\bf 0}_N,{\bf I}_{N})$ is independent of $\theta_x$. ${\bf z} \in \mathbb{C}^{N}$ is a random vector independent of ${\bf x}$ and has zero-mean. For a deterministic matrix ${\bf B} \in \mathbb{C}^{N \times N}$ and given ${\bf C}_y\triangleq\mathbb{E}\left\{{\bf y}{\bf y}^H\right\}$, it holds that
\begin{align}\label{eq:lemma2}
\mathbb{E}\left\{{\bf y}^H{\bf B}{\bf x}\right\}=&\alpha{\bf \bar{x}}^H{\bf B}{\bf \bar{x}}+\alpha\sigma_x^2\tr\left({\bf B}\right), \\
\mathbb{E}\left\{\left|{\bf y}^H{\bf B}{\bf x}\right|^2\right\}=&2\alpha^2\sigma_x^2\Re\left\{{\bf \bar{x}}^H{\bf B}{\bf \bar{x}}\tr\left({\bf B}^H\right)\right\} \nonumber\\&\hspace{-2cm}+\alpha^2\sigma_x^4\left|\tr\left({\bf B}\right)\right|^2+\tr\left({\bf B}\left({\bf \bar{x}}{\bf \bar{x}}^H+\sigma_x^2{\bf I}_N\right){\bf B}^H{\bf C}_y\right). \label{eq:applemma2-claima}
\end{align}

\begin{IEEEproof} Compute $\mathbb{E}\left\{{\bf y}^H{\bf B}{\bf x}\right\}$ as
	\begin{align}
	\mathbb{E}\left\{{\bf y}^H{\bf B}{\bf x}\right\}=&\alpha\mathbb{E}\left\{{\bf x}^H{\bf B}{\bf x}\right\}+\mathbb{E}\left\{{\bf z}^H{\bf B}{\bf x}\right\}\nonumber\\\stackrel{(a)}=&\alpha{\bf \bar{x}}^H{\bf B}{\bf \bar{x}}+\alpha\sigma_x^2\tr\left({\bf B}\right)
	\end{align}
	where we used the independence of zero-mean ${\bf z}$ and ${\bf x}$, and $\mathbb{E}\left\{{\bf x}{\bf x}^H\right\}={\bf \bar{x}}{\bf \bar{x}}^H+\sigma_x^2{\bf I}_N$ in $(a)$.
	
	Let us compute now $\mathbb{E}\left\{\left|{\bf y}^H{\bf B}{\bf x}\right|^2\right\}$ as
	\begin{align}
	&\mathbb{E}\left\{\left|{\bf y}^H{\bf B}{\bf x}\right|^2\right\}=\mathbb{E}\left\{\left(\alpha{\bf x}^H+{\bf z}^H\right){\bf B}{\bf x}{\bf x}^H{\bf B}^H\left(\alpha{\bf x}+{\bf z}\right)\right\}
\nonumber\\	&\stackrel{(a)}=\alpha^2\mathbb{E}\left\{{\bf x}^H{\bf B}{\bf x}{\bf x}^H{\bf B}^H{\bf x}\right\}+\mathbb{E}\left\{{\bf z}^H{\bf B}{\bf x}{\bf x}^H{\bf B}^H{\bf z}\right\} \nonumber\\
&	\stackrel{(b)}=\alpha^2{\bf \bar{x}}^H{\bf B}{\bf \bar{x}}{\bf \bar{x}}^H{\bf B}^H{\bf \bar{x}}+\alpha^2\sigma_x^2\mathbb{E}\left\{{\bf \bar{x}}^H{\bf B}{\bf \bar{x}}{\bf w}^H{\bf B}^H{\bf w}\right\}\nonumber\\&\hspace{0.2cm}+\alpha^2\sigma_x^2\mathbb{E}\left\{{\bf \bar{x}}^H{\bf B}{\bf w}{\bf w}^H{\bf B}^H{\bf \bar{x}}\right\}+\alpha^2\sigma_x^2\mathbb{E}\left\{{\bf w}^H{\bf B}{\bf \bar{x}}{\bf \bar{x}}^H{\bf B}^H{\bf w}\right\}\nonumber\\
	&\hspace{0.2cm}+\alpha^2\sigma_x^2\mathbb{E}\left\{{\bf w}^H{\bf B}{\bf w}{\bf \bar{x}}^H{\bf B}^H{\bf \bar{x}}\right\}+\alpha^2\sigma_x^4\mathbb{E}\left\{{\bf w}^H{\bf B}{\bf w}{\bf w}^H{\bf B}^H{\bf w}\right\}\nonumber\\
	&\hspace{0.2cm}+\tr\left({\bf B}\left({\bf \bar{x}}{\bf \bar{x}}^H+\sigma_x^2{\bf I}_N\right){\bf B}^H\mathbb{E}\left\{{\bf z}{\bf z}^H\right\}\right) \nonumber \\
	&\stackrel{(c)}=\alpha^2{\bf \bar{x}}^H{\bf B}{\bf \bar{x}}{\bf \bar{x}}^H{\bf B}^H{\bf \bar{x}}+\alpha^2\sigma_x^2{\bf \bar{x}}^H{\bf B}{\bf \bar{x}}\tr\left({\bf B}^H\right)\nonumber\\&\hspace{0.2cm}+\alpha^2\sigma_x^2{\bf \bar{x}}^H{\bf B}{\bf B}^H{\bf \bar{x}}+\alpha^2\sigma_x^2{\bf \bar{x}}^H{\bf B}^H{\bf B}{\bf \bar{x}}\nonumber\\&\hspace{0.2cm}+\alpha^2\sigma_x^2{\bf \bar{x}}^H{\bf B}^H{\bf \bar{x}}\tr\left({\bf B}\right)+\alpha^2\sigma_x^4\left|\tr\left({\bf B}\right)\right|^2+\alpha^2\sigma_x^4\tr\left({\bf B}{\bf B}^H\right)\nonumber\\&\hspace{0.2cm}+\tr\left({\bf B}\left({\bf \bar{x}}{\bf \bar{x}}^H+\sigma_x^2{\bf I}_N\right){\bf B}^H\left({\bf C}_y-\alpha^2\left({\bf \bar{x}}{\bf \bar{x}}^H+\sigma_x^2{\bf I}_N\right)\right)\right),\label{eq:applemma2-2b}
	\end{align}
	where we used the independence of zero-mean ${\bf z}$ and ${\bf x}$ in $(a)$ and $(b)$. We have written all the non-zero individual terms of $\mathbb{E}\left\{{\bf x}^H{\bf B}{\bf x}{\bf x}^H{\bf B}^H{\bf x}\right\}$ separately by noting that $\theta_x$ is independent of ${\bf w}$ and ${\bf w}$ is circularly symmetric in $(b)$. We have used the cyclic shift property of trace and Lemma~\ref{applemma1} in $(c)$ together with $\mathbb{E}\left\{{\bf z}{\bf z}^H\right\}={\bf C}_y-\alpha^2\left({\bf \bar{x}}{\bf \bar{x}}^H+\sigma_x^2{\bf I}_N\right)$. After arranging the terms in \eqref{eq:applemma2-2b}, we obtain the result in \eqref{eq:applemma2-claima}.
\end{IEEEproof}
\end{lemma}

\section{Proof of Lemma 1\label{lemma1_proof}}
Let us calculate the average input power with LMMSE for the MR precoder ${\bf w}_{kl}={\bf \hat{g}}_{kl}$ in \eqref{eq:lmmse}. By using Lemma~\ref{applemma2} with ${\bf x}={\bf g}_{kl}$, ${\bf \bar{x}}={\bf \bar{g}}_{kl}$, $\sigma_x^2=\beta_{kl}$, ${\bf y}={\bf z}_{il}$, $\alpha=\sqrt{\tau_p\rho_p}$, ${\bf B}=\sqrt{\tau_p\rho_p}{\bf \Psi}_{il}^{-1}{\bf R}_{il}$, and ${\bf C}_y={\bf \Psi}_{il}$, we obtain
\begin{align}
&\mathbb{E}\left\{{\bf \hat{g}}_{il}^H{\bf g}_{kl}{\bf g}_{kl}^H{\bf \hat{g}}_{il}\right\}=\nonumber\\&\hspace{0.2cm}2\tau_p^2\rho_p^2\beta_{kl}\Re\left\{{\bf \bar{g}}_{kl}^H{\bf \Psi}_{il}^{-1}{\bf R}_{il}{\bf \bar{g}}_{kl}\tr\left({\bf R}_{il}{\bf \Psi}_{il}^{-1}\right)\right\}\nonumber\\&\hspace{0.2cm}+\tau_p^2\rho_p^2\beta_{kl}^2\left|\tr\left({\bf \Psi}_{il}^{-1}{\bf R}_{il}\right)\right|^2+\tau_p\rho_p\tr\left({\bf \Psi}_{il}^{-1}{\bf R}_{il}{\bf R}_{kl}{\bf R}_{il}\right), \nonumber\\& \ i\in \mathcal{P}_k. \label{eq:applemma2-1a}
\end{align}
For the expectation $\mathbb{E}\left\{{\bf \hat{g}}_{il}^H{\bf g}_{kl}{\bf g}_{kl}^H{\bf \hat{g}}_{il}\right\}$ for $i \notin \mathcal{P}_k$, ${\bf \hat{g}}_{il}$ and ${\bf g}_{kl}$ are independent and we have
\begin{align}
\mathbb{E}\left\{{\bf \hat{g}}_{il}^H{\bf g}_{kl}{\bf g}_{kl}^H{\bf \hat{g}}_{il}\right\}=&\tr\left(\mathbb{E}\left\{{\bf \hat{g}}_{il}{\bf \hat{g}}_{il}^H\right\}\mathbb{E}\left\{{\bf g}_{kl}{\bf g}_{kl}^H\right\}\right)\nonumber\\\stackrel{(a)}=&\tr\left({\bf \hat{R}}_{il}{\bf R}_{kl}\right), \ \ \ i \notin \mathcal{P}_k,\label{eq:applemma2-2a}
\end{align}
where we used \eqref{eq:Rhat} in $(a)$. By using Lemma~\ref{applemma2} with ${\bf x}={\bf g}_{kl}$, ${\bf \bar{x}}={\bf \bar{g}}_{kl}$, $\sigma_x^2=\beta_{kl}$, ${\bf y}={\bf z}_{il}$, $\alpha=\sqrt{\tau_p\rho_p}$, ${\bf B}=\sqrt{\tau_p\rho_p}{\bf \Psi}_{il}^{-1}{\bf R}_{il}$, we obtain
\begin{align}
&\mathbb{E}\left\{{\bf \hat{g}}_{il}^H{\bf g}_{kl}\right\}=\tau_p\rho_p{\bf \bar{g}}_{kl}^H{\bf \Psi}_{il}^{-1}{\bf R}_{il}{\bf \bar{g}}_{kl}+\tau_p\rho_p\beta_{kl}\tr\left({\bf \Psi}_{il}^{-1}{\bf R}_{il}\right), \nonumber\\ & i \in \mathcal{P}_{k}. \label{eq:applemma2-3a}
\end{align}
The expectation $\mathbb{E}\left\{{\bf \hat{g}}_{il}^H{\bf g}_{kl}\right\}=0$ for $i \notin \mathcal{P}_k$ since ${\bf \hat{g}}_{il}$ and ${\bf g}_{kl}$ are independent and have zero mean. Let us now evaluate the expectations for the MR precoder ${\bf w}_{kl}={\bf z}_{kl}$ in \eqref{eq:suff-stats} for the LS-based channel estimation. By using Lemma~\ref{applemma2} with ${\bf x}={\bf g}_{kl}$, ${\bf \bar{x}}={\bf \bar{g}}_{kl}$, $\sigma_x^2=\beta_{kl}$, ${\bf y}={\bf z}_{il}$, $\alpha=\sqrt{\tau_p\rho_p}$, ${\bf B}={\bf I}_N$, and ${\bf C}_y={\bf \Psi}_{il}$, we obtain
\begin{align}
\mathbb{E}\left\{{\bf z}_{il}^H{\bf g}_{kl}{\bf g}_{kl}^H{\bf z}_{il}\right\}=&2N\tau_p\rho_p\beta_{kl}{\bf \bar{g}}_{kl}^H{\bf \bar{g}}_{kl}+N^2\tau_p\rho_p\beta_{kl}^2\nonumber\\&+\tr\left({\bf R}_{kl}{\bf \Psi}_{il}\right), \ \ i\in \mathcal{P}_k.\label{eq:applemma2-1b}
\end{align}
For the expectation $\mathbb{E}\left\{{\bf z}_{il}^H{\bf g}_{kl}{\bf g}_{kl}^H{\bf z}_{il}\right\}$ for $i \notin \mathcal{P}_k$, ${\bf z}_{il}$ and ${\bf g}_{kl}$ are independent and we have
\begin{align}
\mathbb{E}\left\{{\bf z}_{il}^H{\bf g}_{kl}{\bf g}_{kl}^H{\bf z}_{il}\right\}=&\tr\left(\mathbb{E}\left\{{\bf z}_{il}{\bf z}_{il}^H\right\}\mathbb{E}\left\{{\bf g}_{kl}{\bf g}_{kl}^H\right\}\right)\nonumber\\&=\tr\left({\bf \Psi}_{il}{\bf R}_{kl}\right), \ \ \ i \notin \mathcal{P}_k. \label{eq:applemma2-2bb}
\end{align}
By using Lemma~\ref{applemma2} with ${\bf x}={\bf g}_{kl}$, ${\bf \bar{x}}={\bf \bar{g}}_{kl}$, $\sigma_x^2=\beta_{kl}$, ${\bf y}={\bf z}_{il}$, $\alpha=\sqrt{\tau_p\rho_p}$, ${\bf B}={\bf I}_N$, we obtain
\begin{align}
&\mathbb{E}\left\{{\bf z}_{il}^H{\bf g}_{kl}\right\}=\sqrt{\tau_p\rho_p}{\bf \bar{g}}_{kl}^H{\bf \bar{g}}_{kl}+N\sqrt{\tau_p\rho_p}\beta_{kl}, \ \ i \in \mathcal{P}_{k}. \label{eq:applemma2-3b}
\end{align}
The expectation $\mathbb{E}\left\{{\bf z}_{il}^H{\bf g}_{kl}\right\}=0$ for $i \notin \mathcal{P}_k$ since ${\bf z}_{il}$ and ${\bf g}_{kl}$ are independent and have zero mean. If we insert the expectations calculated above into \eqref{eq:Ik}, we obtain the results in \eqref{eq:Ik-lemma1} and \eqref{eq:Ik-lemma2}.

\section{Proof of Lemma 2\label{lemma2_proof}}
Let us compute the expectations in the claim of Lemma 2 for the MR decoder ${\bf v}_{kl}={\bf \hat{g}}_{kl}$ in \eqref{eq:lmmse}. By using the result in \eqref{eq:applemma2-3a}, we obtain
\begin{align}
b_{kl}&=\mathbb{E}\left\{{\bf \hat{g}}_{kl}^H{\bf g}_{kl}\right\}\nonumber\\&=\tau_p\rho_p{\bf \bar{g}}_{kl}^H{\bf \Psi}_{kl}^{-1}{\bf R}_{kl}{\bf \bar{g}}_{kl}+\tau_p\rho_p\beta_{kl}\tr\left({\bf \Psi}_{kl}^{-1}{\bf R}_{kl}\right).
\end{align}
By using the results in \eqref{eq:applemma2-1a} and  \eqref{eq:applemma2-2a} we have 
\begin{align}
c_{ki}^{ll}=&\mathbb{E}\left\{{\bf \hat{g}}_{kl}^H{\bf g}_{il}{\bf g}_{il}^H{\bf \hat{g}}_{kl}\right\}\nonumber\\=&2\tau_p^2\rho_p^2\beta_{il}\Re\left\{{\bf \bar{g}}_{il}^H{\bf \Psi}_{kl}^{-1}{\bf R}_{kl}{\bf \bar{g}}_{il}\tr\left({\bf R}_{kl}{\bf \Psi}_{kl}^{-1}\right)\right\}\nonumber\\&+\tau_p^2\rho_p^2\beta_{il}^2\left|\tr\left({\bf \Psi}_{kl}^{-1}{\bf R}_{kl}\right)\right|^2\nonumber\\
&+\tau_p\rho_p\tr\left({\bf \Psi}_{kl}^{-1}{\bf R}_{kl}{\bf R}_{il}{\bf R}_{kl}\right), \ \ i\in \mathcal{P}_k. \\
c_{ki}^{ll}=&\tr\left({\bf \hat{R}}_{kl}{\bf R}_{il}\right), \ \ \ i \notin \mathcal{P}_k.
\end{align}
The expectation $c_{ki}^{ll^{\prime}}=\mathbb{E}\left\{{\bf \hat{g}}_{kl}^H{\bf g}_{il}{\bf g}_{il^{\prime}}^H{\bf \hat{g}}_{kl^{\prime}}\right\}$ for $l^{\prime}\neq l$ is given by using \eqref{eq:applemma2-3a} as
\begin{align}
&c_{ki}^{ll^{\prime}}=\tau_p^2\rho_p^2\Big({\bf \bar{g}}_{il}^H{\bf \Psi}_{kl}^{-1}{\bf R}_{kl}{\bf \bar{g}}_{il}+\beta_{il}\tr\left({\bf \Psi}_{kl}^{-1}{\bf R}_{kl}\right)\Big)\times\nonumber\\&\hspace{0.2cm}\Big({\bf \bar{g}}_{il^{\prime}}^H{\bf \Psi}_{kl^{\prime}}^{-1}{\bf R}_{kl^{\prime}}{\bf \bar{g}}_{il^{\prime}}+\beta_{il^{\prime}}\tr\left({\bf \Psi}_{kl^{\prime}}^{-1}{\bf R}_{kl^{\prime}}\right)\Big)^* , \ i\in \mathcal{P}_k,  \   l^{\prime}\neq l, \nonumber\\&  c_{ki}^{ll^{\prime}}=0, \ \ i\notin \mathcal{P}_k, \  \  l^{\prime}\neq l.
\end{align}
The expectation $d_{kl}=\mathbb{E}\left\{ {\bf \hat{g}}_{kl}^H{\bf n}_l^I\left({\bf n}_l^I\right)^H{\bf \hat{g}}_{kl}\right\}$ is given by $d_{kl}=\sigma^2\tr\left({\bf \hat{R}}_{kl}\right)$ using the independence of data noise and the channel estimate.

\section{Proof of Lemma 4\label{lemma4proof}}
Let $R$ denote the rank of $\bm{P}_k^{\star}$ for some $k$. In this case, $\bm{P}_k$ can be expressed as
\begin{align}
\bm{P}_k^{\star}=\sum_{r=1}^R\bm{x}_r\bm{x}_r^T,
\end{align}
for some vectors $\bm{x}_r\in \mathbb{R}^{L}$. Let us now consider another $\bm{P}_k^{\star\star}$ with rank one and equal diagonal elements with $\bm{P}_k^{\star}$ in \eqref{eq:Pk}. Since $\bm{P}_k^{\star\star}$ has rank one, it can be expressed as $\bm{P}_k^{\star\star}=\bm{y}\bm{y}^T$ for some vector $\bm{y} \in \mathbb{R}^{L}$ such that
\begin{align}\label{eq:appe}
\sum_{r=1}^Rx_{rl}^2=y_l^2, \ \ \ l=1,\ldots,L,
\end{align}
where $x_{rl}$ and $y_l$ denote the $l^{\textrm{th}}$ element of the vectors $\bm{x}_r$ and $\bm{y}$, respectively. The equality in \eqref{eq:appe} follows from the diagonal elements of $\bm{P}_k^{\star}$ and $\bm{P}_k^{\star\star}$ being the same. Using Cauchy-Schwarz inequality, it can be shown that
\begin{align}
&{P_k^{ll^{\prime}}}^{\star}=\sum_{r=1}^Rx_{rl}x_{rl^{\prime}}\leq \sqrt{\sum_{r=1}^Rx_{rl}^2}\sqrt{\sum_{r=1}^Rx_{rl^{\prime}}^2}=y_ly_{l^{\prime}}={P_k^{ll^{\prime}}}^{\star\star},\nonumber\\&
l\neq l^{\prime}.
\end{align}
Hence, all the off-diagonal elements of $\bm{P}_k^{\star\star}$ and the harvested energy for each UE is larger when $\bm{P}_k^{\star}$ is replaced by  $\bm{P}_k^{\star\star}$ without affecting other constraints and the objective function. Hence, this new solution with $\bm{P}_k^{\star\star}$ in \eqref{eq:Pk} is also optimum to the considered problem. 

\ifCLASSOPTIONcaptionsoff
  \newpage
\fi

\end{document}